\documentclass[11pt]{article}
\usepackage{amsmath}
\usepackage{amssymb}
\usepackage{latexsym}
\usepackage{epsfig}

\newcommand{\be}{\begin{equation}}
\newcommand{\ee}{\end{equation}}

\begin{document}

{}~\hfill\vbox{\hbox{CTP-SCU/2014006}\hbox{CAS-KITPC/ITP-444}
}\break

\vskip 1.0cm

\centerline{\Large \bf Effects of quantum gravity on black holes}

\vspace*{12.0ex}

\centerline{\large Deyou Chen$^{1,2}$, Houwen Wu$^1$, Haitang Yang$^{1,3}$ and Shuzheng Yang$^2$}

\vspace*{7.0ex}

\vspace*{4.0ex}

\centerline{\large \it $^{1}$Center for Theoretical Physics,}

\centerline{\large \it College of Physical Science and Technology,}

\centerline{\large \it Sichuan University, Chengdu 610064, China}
\vspace*{2.0ex}

\centerline{\large \it $^{2}$College of Physics and Electronic Information,}

\centerline{\large \it China West Normal University,}

\centerline{\large \it Nanchong 637009, China}
\vspace*{2.0ex}

\centerline{\large \it $^{3}$Kavli Institute for Theoretical Physics China (KITPC),}

\centerline{\large \it Chinese Academy of Sciences,}

\centerline{\large \it Beijing 100080, P.R. China}
\vspace*{2.0ex}

\centerline{dchen@cwnu.edu.cn, 2013222020003@stu.scu.edu.cn, hyanga@scu.edu.cn, szyang@cwnu.edu.cn}

\vspace*{10.0ex}

\centerline{\bf Abstract}
\bigskip
\smallskip

In this review, we discuss effects of quantum gravity on black hole physics. After a brief review of the origin of the minimal observable length from various quantum gravity theories, we present the tunneling method.  To incorporate quantum gravity effects, we modify the Klein-Gordon equation and Dirac equation by the modified fundamental commutation relations. Then we use the modified equations to discuss the tunneling radiation of scalar particles and fermions. The corrected Hawking temperatures are related to the quantum numbers of the emitted particles. Quantum gravity corrections slow down the increase of the temperatures. The remnants are observed as $M_{\hbox{Res}}\gtrsim \frac{M_p}{\sqrt{\beta_0}}$. The mass is quantized by the modified Wheeler-DeWitt equation and is proportional to $n$ in quantum gravity regime. The thermodynamical property of the black hole is studied by the influence of quantum gravity effects.
\vspace*{1.5ex}


\vspace*{1.0ex}


\vfill \eject

\baselineskip=16pt

\vspace*{10.0ex}

\tableofcontents

\section{Introduction}

The discovery of Hawking radiation is an important progress in gravitational physics \cite{SWH}. It shows the existence of the thermodynamical property in black holes. The temperature of a black hole is proportional to its surface gravity. The entropy is equivalent to one-forth of the horizon area. These results reveal the connection between relativity theory, statistical mechanics and quantum mechanics.

Hawking radiation is described as a tunneling effect of particles across the black hole's horizon. It has been deeply researched by different methods \cite{DR,SS,KW1,KW2,PW,SP}. The original research was carried out in the quantum field theory which is based on the Heisenberg uncertainty principle. It led to the standard Hawking formula and predicted the complete evaporation of black holes.

The tunneling method put forward by Kraus and Wilczek is an effective way to study the Hawking radiation and attracts many physicists attention \cite{KW1,KW2}. In this method, the derivation of Hawking radiation is based on the dynamical geometry approach and relies on the calculation of the imaginary part of the classically forbidden process \cite{PW,PW1}. When energy conservation is taken into account, the black hole's spacetime should be varied with the emission of a particle. Considering the varied spacetime and the self-gravity interaction, Parikh and Wilczek researched the tunneling radiation of massless particles across the horizons of the spherically symmetric black holes. The result showed that tunneling rate is related to the change of Bekenstein-Hawking entropy. The corrected Hawking temperature is higher than the standard one and the leading correction is dependent on the emitted particle's energy. This result implies that the varied background spacetime accelerates the evaporation of the black hole. In this method, the key point is the derivation of the radial geodesic equation. Therefore, this method is also called as the null geodesic method. For the massless particle, its radial geodesic equation is derived by $ds^2=0$. The equation of motion of a massive particle is different from that of massless case. The trajectory of a massless particle is a null geodesic, while that of a massive particle obeys de Broglie wave function and is the phase velocity of the outgoing wave. The phase velocity is related to the the group velocity. Through solving the group velocity, Zhang and Zhao got the equation of motion of the massive particle \cite{ZZ1,ZZ2}. Then using the equation of motion, the tunneling radiation of massive particles was studied. The result is full in consistence with that derived in Ref. \cite{PW}. Extending this method to the general spacetimes, the tunneling radiation of the various spacetimes were discussed in Refs. \cite{JWC,JW,ECV1,HKV,AJM,ECV2,ZZ3,WJ,KIM,TP,SK,Mitra,BM,BK}.

Another way to derive the action relies on solving the Hamilton-Jacobi equation, which is called as the Hamilton-Jacobi method \cite{ANVZ,KM1}. This method is a extension of the complex path integral method \cite{SP,SPS,SSP}. To derive the imaginary part, the action is separated by the radial and angular parts. In the calculation, it is not needed to assume a particle with the s-wave emission since the contribution of the angular part is canceled out. Using the Hamilton-Jacobi method and the Dirac equation in curved spacetime, the fermions' tunneling behaviors were discussed in Kerner and Mann's work \cite{KM2,KM3}. Due to the fixed background spacetimes, the standard Hawking formulae were recovered. The subsequent work in the virous spacetimes can be found in Refs. \cite{BM1,BM2,SVZ,CV,LRW,CJZ,HCNVS,CJZ1,JIANG,ZY,AS,RS,CYZ,LY1}.

The black hole entropy has been deeply studied. In the original derivation, the entropy is proportional to the horizon area. However, there is a logarithmic correction to the Bekenstein-Hawking entropy in other investigations \cite{KM111,F111,B111}. This phenomenon may be caused by quantum gravity. An interest prediction of quantum gravity theory is the existence of the minimal observable length. This prediction was found in string theory, loop quantum gravity and doubly special relativity, respectively \cite{PKT,ACV,LJG,GAC}. This view was also supported by the Gedanken experiments in the spirit of black hole physics \cite{FS111}. The generalized uncertainty principle (GUP) is an effective way to realize this length,

\begin{eqnarray}
\Delta x \Delta p \geq \frac{\hbar}{2}\left[1+ \beta (\Delta p)^2+ \beta \left\langle p \right\rangle^2\right],
\label{eq1.1}
\end{eqnarray}

\noindent where $\beta=\beta_{0}\ell_{p}^{2}/\hbar^{2}=\beta_{0}c^{2}/M_{p}^{2}$, $\beta_0 <10^{5}$ is a dimensionless parameter \cite{LMS}, $\ell_{p}=\sqrt{G\hbar/c^{3}}$ is the Planck mass and $M_{p}=\sqrt{\hbar c/G}$ is the Planck mass. GUP was first derived in Kempf and Mann's work by the modified fundamental commutation relations  $\left[x_i,p_j\right]= i \hbar \delta_{ij}\left[1+ \beta p^2\right]$, where $x_i$ and $p_i$ are operators of position and momentum defined by \cite{KMM}

\begin{eqnarray}
x_i &=& x_{0i}, \nonumber\\
p_i &=& p_{0i} (1 + \beta p_0^2),
\label{eq1.2}
\end{eqnarray}

\noindent $p_0^2 = \sum p_{0j}p_{0j}$, $x_{0i}$ and $p_{0j}$ satisfy the canonical commutation relations $\left[x_{0i},p_{0j}\right]= i \hbar \delta_{ij}$. Thus the minimal position uncertainty is

\begin{eqnarray}
\Delta x &=& \hbar \sqrt{\beta} \sqrt{1+ \beta \left\langle p \right\rangle^2} .
\label{eq1.3}
\end{eqnarray}

\noindent It implies the minimum measurable length $ \Delta x_0 = \hbar \sqrt{\beta}$. When $\beta >0$ , $\Delta x_0 $ has the physical meaning. Modifications of the commutation relations are not unique. Other modifications can be found in the related references \cite{FB,ADV,DV}.

These modifications play an important role in the researches of quantum gravity. Taking into account effects of quantum gravity, the Schrodinger equation, Klein-Gordon equations and Dirac equation were modified by the modified fundamental commutation relations \cite{NK,HBHRSS}. Applying these modifications to the superconductivity and  quantum Hall effect, the authors computed the Planck scale corrections and showed that Planck scale effects may account for a part of the anomalous magnetic moment of the muon \cite{DM}. The effects of quantum gravity on compact star cores and on the Planck era of the universe were discussed in the papers \cite{WYZ1,WYZ2,BDV,CDAV}. The treatment of hydrogen atom problems with a generalized uncertainty relation was present in the work of Bouaziz et al. \cite{BF,FAFN,ST}. Incorporating the GUP into black holes, the thermodynamical properties and the remnants were studied in recent researches \cite{ACS,BG,SA,LW,MAJ,ZDM,NS}. The corrections of quantum gravity to the various quantum phenomena appeared \cite{DV}.

In this paper, we review quantum gravity effects on the black holes. We first give a review of the minimal observable length derived in various quantum gravity theories in section 2. Then, the tunneling method was discussed in section 3. This method contains the null geodesic method and the Hamilton-Jacobi method. In section 4, we adopt the new definition for the geodesic equation of a massive particle to discuss the tunneling radiation. In section 5, incorporating effects of quantum gravity into black holes' physics, we investigate the tunneling radiation of scalar particles and fermions. Effects of quantum gravity were introduced by the modified Klein-Gordon equation and Dirac equation. The corrected temperatures are affected by the quantum numbers of the emitted particles. Quantum corrections slow down the increase of the temperatures. The remnants are left during the evaporation. In Section 6, we first review the quantized mass of the black hole by the Wheeler-DeWitt (WDW) equation. Then modify WDW equation by the influence of the quantum gravity effects and use the modified equation to discuss the quantum black hole. The thermodynamics of the black holes is investigated in section 7. Section 8 is devoted to our discussion and conclusion.

\section{The minimal observable length}

In this section, we do not review the history of the minimal observable length, since it can be seen in the review paper \cite{Hossenfelder:2012jw}. Here we only present the motivation which inspired GUP.

\subsection{The viewpoint of mini black holes}

There are two critical scales in the modern physics. The first one is the Compton wavelength which marks the quantum property. The second one is the radius of a black hole. In this scale, there is a strong gravitational field. When the mass of the black hole approaches to the Planck mass, the radius and its Compton wavelength are comparable \cite{Hossenfelder:2012jw}. Thus we should consider the effects of quantum and strong gravitation together in this region. We first take into account the Compton wavelength, which is defined by

\begin{equation}
\lambda=\frac{1}{m},
\end{equation}

\noindent where $\hbar=c$ are set to $1$ and $m$ is the rest mass of a given particle. It can be seen as the wavelength of a photon. Moreover, it is also the limitation measured a particle's position \cite{Garay:1994en}. To see this point we use the Heisenberg uncertainty relation

\begin{equation}
\triangle x\triangle p\geq\frac{1}{2}.
\end{equation}

\noindent For a relativistic particle, we adopt $E\sim p\sim m$. If $\triangle x$ is smaller than the Compton wavelength, namely $\triangle x\leq\frac{1}{E}$,
we will find $\triangle p\geq\frac{1}{2}E$. This will create other particle and lead to the definition of a particle unclear. Therefore, the position uncertainty must satisfy the following inequality

\begin{equation}
\triangle x\geq\frac{1}{2m},
\end{equation}

\noindent which is larger than the Compton wavelength.

On the other hand, the radius of a Schwarzschild black hole is

\begin{equation}
r_{H}=2GM,
\end{equation}

\noindent where $M$ is a mass of black hole. The uncertainty for the radius can be directly found as

\begin{equation}
\triangle r_{H}=2G\triangle M.
\end{equation}

\noindent Thus, we get the uncertainty of the position is longer than the Compton wavelength, namely, $\triangle x\geq\frac{1}{2}\frac{1}{\triangle M}$. When $\triangle M$ of a black hole approaches to the Planck mass or the radius reaches its Compton wavelength, the two terms become comparable $\frac{1}{\triangle M}\sim G\triangle M$, since $M_{p}=G^{-1/2}$. Therefore, the total uncertainty relation should be combined by these two contributions linearly

\begin{equation}
\triangle r_{H}\geq\frac{1}{2}\left(\frac{1}{\triangle M}+\beta_{0}G\triangle M\right),
\end{equation}

\noindent where $\beta_{0}$ is a dimensionless constant to specify a certain theory. Therefore, the minimal length appears in the above equation.

\subsection{The viewpoint of string theory}

We first begin with a non-relativistic particle. The Feynman path integral of this particle can be achieved by considering its world line $x\left(\tau\right)$ \cite{Garay:1994en,Montani:2011zz} , which is given as follows

\begin{equation}
Z_{\mathrm{particle}}\sim\int\mathcal{D}x\exp\left[\frac{i}{\lambda^{2}}\sum\epsilon\left(\frac{\delta_{\tau}x}{\epsilon}\right)^{2}\right],
\end{equation}

\noindent where $\lambda$ is a dimensional constant and $\epsilon$ is a parameter which comes from that we divide a time interval into the equal pieces. For a piece, we find that the particle travels a distance

\begin{equation}
\left\langle \left(\delta x\right)^{2}\right\rangle \sim\epsilon\lambda^{2}.
\end{equation}

\noindent It implies that the resolution of the above equation can be arbitrarily high when $\epsilon\rightarrow0$. Now we back to the string theory and introduce a fundamental action in string theory. It is the well-known Polyakov action, which describes a dynamics on the world sheet in a target space. The action is

\begin{equation}
S=-\frac{1}{4\pi\alpha^{\prime}}\int d^{2}\sigma\sqrt{-g}g^{\alpha\beta}\partial_{\alpha}X^{\mu}\partial_{\beta}X^{\nu}\eta_{\mu\nu},
\end{equation}

\noindent where $g_{\alpha\beta}$ is a metric on the worldsheet and $X^{\mu}\left(\tau,\sigma\right)$ is a surface which defines a map from the worldsheet to the target space. It should be noted that the prefactor of this action denotes the fundamental length of a string:

\begin{equation}
\ell_{s}^{2}=\alpha^{\prime},
\end{equation}

\noindent where the Planck scale in $10$ dimensions has $\ell_{p}^{\left(10\right)}\sim g^{1/4}\ell_{s}$, with the string coupling $g\sim e^{\phi}$. It implies that if the string coupling $g$ is small enough, the string length is larger than the Planck length in the given dimensions. Now, we can get the path integral for this action as\cite{Garay:1994en,Montani:2011zz}

\begin{equation}
Z_{\mathrm{string}}\sim\int\mathcal{D}X\exp\left(\frac{i}{\ell_{s}^{2}}\sum\epsilon^{2}\left[\left(\frac{\delta_{\tau}X}{\epsilon}\right)^{2}+
\left(\frac{\delta_{\sigma}X}{\epsilon}\right)^{2}\right]\right).
\end{equation}

\noindent The resolution becomes

\begin{equation}
\left\langle \left(\delta X\right)^{2}\right\rangle \sim\ell_{s}^{2}.
\end{equation}

\noindent This formula is independent of the parameter $\epsilon$, which implies that no matter how the parameter $\epsilon$ is, there is a minimum length $\ell_{s}$. Moreover, when we consider the scattering amplitude of a string at high energy, it also gives us the same result. Since, the string theory is fuzzy below the scale $\ell_{s}=\sqrt{\alpha^{\prime}}$, we can not use strings to probe the distance shorter than this scale \cite{Tong:2009np}.

The minimal length can be also seen in the T-duality of a closed string. We consider a closed string winding on the compact extra dimension. In this configuration, the closed string can not shrink to zero size. It is different from the closed string which moves in the Minkowski spacetime and gives us a remarkable property. In the $1+25$ dimensional spacetime, we first compact the $25$th extra dimension, which needs us to identify the coordinate of $25$th dimension via

\begin{equation}
x\sim x+2\pi R,
\end{equation}

\noindent where $R$ is a radius of the extra dimension. In this compact background, the periodicity condition of the closed string $X\left(\tau,\sigma+2\pi\right)=X\left(\tau,\sigma\right)$ can be replaced by

\begin{equation}
X\left(\tau,\sigma+2\pi\right)=X\left(\tau,\sigma\right)+m\left(2\pi R\right),
\end{equation}

\noindent where $m$ is called as the winding number. For convenience, we introduce the definition of the winding: $w\equiv mR/\alpha^{\prime}$, which is born from the winding number. Therefore, the mode expansions for the left-moving and right-moving string coordinates are

\begin{eqnarray*}
X_{L}\left(\tau+\sigma\right) & = & \frac{1}{2}x_{0}^{L}+\frac{\alpha^{\prime}}{2}\left(p+w\right)\left(\tau+\sigma\right)+\mathrm{oscillator\; modes},\\
X_{R}\left(\tau-\sigma\right) & = & \frac{1}{2}x_{0}^{R}+\frac{\alpha^{\prime}}{2}\left(p-w\right)\left(\tau+\sigma\right)+\mathrm{oscillator\; modes.}
\end{eqnarray*}

\noindent Combining these two coordinates, we find the full coordinates $X\left(\tau,\sigma\right)$, which take a form as

\begin{equation}
X\left(\tau,\sigma\right)=x_{0}+\alpha^{\prime}p\tau+\alpha^{\prime}w\sigma+\mathrm{oscillator\; modes.}
\end{equation}

\noindent It is easy to see that the winding $w$ has a same behavior with the momenta $p$. This is why we call it as winding momenta. To quantize this closed string, the spectrum of $p$ around the circle direction can be obtained directly from $e^{ipX}$ which looks like the unity operator. The spectrum is

\begin{equation}
p=\frac{n}{R},\quad n\in\mathbb{Z}.
\end{equation}

\noindent In view of the relationship between the winding and winding number, we find the eigenvalues of $w$, which is

\begin{equation}
w=\frac{mR}{\alpha^{\prime}},\quad m\in\mathbb{Z}.
\end{equation}

\noindent After calculations, the formula of the squared mass is gotten as

\begin{equation}
M^{2}=p^{2}+w^{2}+\frac{2}{\alpha^{\prime}}\left(N+\tilde{N}-2\right)=\frac{n^{2}}{R^{2}}+\frac{m^{2}R^{2}}{\alpha^{\prime2}}+
\frac{2}{\alpha^{\prime}}\left(N+\tilde{N}-2\right),
\end{equation}

\noindent where $N$ and $\tilde{N}$ are integers and they satisfy
the level matching condition: $N-\tilde{N}=nm$. It is easy to figure
out that this formula is invariant under the exchange

\begin{equation}
R\longleftrightarrow\frac{\alpha^{\prime}}{R},\quad m\longleftrightarrow n.
\end{equation}

\noindent This remarkable statement is named as T-duality in the closed string theory, and it tells us we can not distinguish the physics at very large or very small circle. It also exhibits a minimal length scale in string theory. Since, when a circle shrinks into the size of a critical scale $R=\sqrt{\alpha^{\prime}}$, the theory will become one which is used to describe the physics in a growing circle. In other words, the modes of momenta will be replaced by winding modes. Therefore, it gives us the minimal length \cite{Tong:2009np},

\begin{equation}
R_{\mathrm{min}}=\sqrt{\alpha^{\prime}}=\ell_{s},
\end{equation}

\noindent where $\ell_{s}$ is a string length.

\subsection{The viewpoint of loop quantum gravity}

The discussion of this section is based on Refs. \cite{Gambini:2011zz,Rovelli:2004tv}. Loop quantum gravity (LQG) is a candidate of quantum gravity theory. Unlike string theory, LQG does not need the supersymmetry and extra dimensions. It is only focused on quantizing gravity, but not other gauge fields. However, the motivation of LQG was born from gauge theories. Recall the Maxwell theory, the Stokes' theorem tells us

\begin{equation}
\int_{C}\vec{A}\cdot\vec{ds}=\int_{S}\vec{\nabla}\times\vec{A}\cdot\vec{n}d^{2}x,
\end{equation}

\noindent where LHS describes the circulation of a vector potential $\vec{A}$ along a closed curve $C$, and RHS is the curl of $\vec{A}$ on the surface bounded by the closed curve. This equality tells us that if we find a circulation for all closed curves, we will find the curl of $\vec{A}$. Here,  $\epsilon^{abc} F_{bc} = 2\left(\vec{\nabla}\times\vec{A}\right)^{a}$. The result shows that if we find the circulation for all closed curves, we can know all characteristics of Maxwell fields. This result can be naturally extended to the Yang-Mills theory. Since the Yang-Mills theory is controlled by non-Abelian groups, the Stokes' theorem becomes invalid and we should adopt non-Abelian Stokes' theorem. Roughly speaking, we can use the holonomy to replace the circulation in the Maxwell theory. The result is same to that in the Maxwell theory. If we know the trace of the holonomy for a vector potential $\mathbf{A}_{\mu}=\sum A_{\mu}^{i}\sigma^{i}$ (can be seen as a connection in the Yang-Mills theory) along all loops, we can get all qualities of Yang-Mills fields. In other words, the trace of the holonomy can be used to construct a basis for all observable quantities in the Yang-Mills theory. In view of the new variables introduced by Ashtekar, the general relativity can be seen as the Yang-Mills theory in loop quantum gravity.

Now let us see how a minimal length emerges from LQG. We quantize gravity in the general relativity. The canonical quantization is adopted. Firstly, the Einstein-Hilbert action is rewritten in the Hamiltonian formulation. Secondly, the configuration variables and its conjugate momenta are expressed as operators. Then we impose the commutation relations. Since the space and time are treated as same footing in the general relativity, this does not work in the Hamiltonian formulation. We should carry out decomposition on the spacetime at first. In the decomposition, we assume the spacetime topology is $\varSigma\times R$, where $\varSigma$ is a three dimensional manifold. When the time $t$ is fixed, $\varSigma$ can be seen as a spatial slice of the spacetime. To use triads formulation, the spatial metric on $\varSigma$ is defined by the inner product between three vectors, namely,

\begin{equation}
q^{ab}=E_{i}^{a}E_{j}^{b}\delta^{ij},
\end{equation}

\noindent where $E_{i}^{a}$ is triads, $i=1,2,3$ denote the three-dimensional vector fields, $a$ denotes coordinates of a curved space and $i$ denotes coordinates of at space. The covariant derivative is $D_{a}G^{i}=\partial_{a}G^{i}+\omega_{aj}^{i}G^{j}$, where $\omega$ is a spin connection. To write the Einstein-Hilbert action in terms of triads, we should introduce the concept of triads density, $\tilde{E}_{i}^{a}=\sqrt{\det\left(q\right)}E_{i}^{a}$. Therefore, the spacetime metric can be written as

\begin{equation}
ds^{2}=\left(-N^{2}+N_{a}N^{a}\right)dt^{2}+2N_{a}dtdx^{a}+q_{ab}dx^{a}dx^{b}.\label{eq:foli metric}
\end{equation}

\noindent Now let us recall the Einstein-Hilbert action in the general relativity, which is expressed as follows

\begin{equation}
S=\frac{1}{16\pi G}\int d^{4}x\sqrt{-\det\left(g\right)}R.
\end{equation}

\noindent Using the Ashtekar's variables, we rewrite the action into the canonical formulation. The Ashtekar's variables are defined by

\begin{eqnarray*}
\mathrm{Configuration\; variable}: & A_{a}^{i}, & SU\left(2\right)\;\mathrm{Yang-Mills\; like\; connection},\\
\mathrm{Conjugate\; momenta}: & \tilde{E}_{i}^{a}, & \mathrm{triads\; density\; in\; GR}.
\end{eqnarray*}

\noindent The Poission bracket for Ashtekar's variables is

\begin{equation}
\left\{ A_{a}^{i}\left(x\right),\tilde{E}_{i}^{a}\left(y\right)\right\} =8\pi G\gamma\delta_{b}^{a}\delta_{j}^{i}\delta^{3}\left(x-y\right),
\end{equation}

\noindent where $\gamma$ is the Barbero-Immirzi parameter. Adopting the Ashtekar's variables and using the metric (\ref{eq:foli metric}), the Einstein-Hilbert action is rewritten as

\begin{equation}
S=\frac{1}{8\pi G\gamma}\int dt\int_{\Sigma}d^{3}x\left[\tilde{E}_{i}^{a}\dot{A}_{a}^{i}+NH+N^{a}H_{a}+\lambda_{i}G^{i}\right],
\end{equation}

\noindent with

\begin{eqnarray}
H & = & \epsilon_{ijk}\tilde{E}_{i}^{a}\tilde{E}_{j}^{b}F_{ab}^{k},\nonumber \\
H_{a} & = & \tilde{E}_{i}^{b}F_{ab}^{i},\nonumber \\
G^{i} & = & D_{a}\tilde{E}_{i}^{a},\label{eq:constraint}
\end{eqnarray}

\noindent where $F_{ab}$ is a field strength of the connection $A_{a}^{i}$ and $N$, $N_{a}$, $\lambda_{i}$ are Lagrange multipliers. Since the general relativity is a totally constrained system, the Hamiltonian of the action equals zero and it is only a linear combination of some constraints. In Eq. (\ref{eq:constraint}), $H=0$ is a Hamiltonian constraint, $H_{a}=0$ is a momentum constraint and $G^{i}=0$ is a Gauss' law. Keep in mind that Gauss' law generates gauge transformations, and the last two constraints of (\ref{eq:constraint}) manifest diffeomorphisms. In the quantization, these constraints control the behavior of the wave function $\Psi$ and the dynamics of the system. We should note that the Hamiltonian constraint is not well-defined. The related discussions is referred to Refs.\cite{Gambini:2011zz,Rovelli:2004tv}. In the following, we use the constraints to pick out the physical states.

To quantize this system, we adopt the procedure of the canonical quantization. After that the canonical variables is written to operators, the commutation relations are given by

\begin{equation}
\left[\hat{A}_{a}^{i}\left(x\right),\hat{\tilde{E}}_{i}^{a}\left(y\right)\right]=i\delta_{b}^{a}\delta_{j}^{i}\delta^{3}\left(x-y\right).
\end{equation}

\noindent The natural wave function is based on the connection $\Psi\left(A_{a}^{i}\right)$. We get

\noindent
\begin{eqnarray}
\hat{A}_{a}^{i}\Psi\left(A\right) & = & A_{a}^{i}\Psi\left(A\right),\nonumber \\
\hat{\tilde{E}}_{i}^{a}\Psi\left(A\right) & = & -i\frac{\delta\Psi\left(A\right)}{\delta A_{a}^{i}}.\label{eq:triad operator}
\end{eqnarray}

\noindent In order to satisfy the constraints, we use the holonomy to construct the wave function. The trace of the holonomy is a function of the connection and it is invariant under gauge transformations. Therefore, we expand the connection representation in a basis of the loop representation,

\begin{equation}
\mathrm{loop\; transformation}:\quad\Psi\left[A\right]=\underset{\gamma}{\sum}\Psi\left[\gamma\right]W_{\gamma}\left[A\right],
\end{equation}

\noindent where $\Psi\left[\gamma\right]$ represents the loop basis and $W_{\gamma}\left[A\right]$ are the traces of the holonomies defined by

\begin{equation}
W_{\gamma}\left[A\right]=\mathrm{Tr}\left(P\left[\exp\left(-\oint_{\gamma}\dot{\gamma}^{a}\left(s\right)\mathbf{A}_{a}\left(s\right)\right)ds\right]\right),
\end{equation}

\noindent with a path ordered exponential $P$. There is a problem for the loop basis: it is over complete. To solve this problem, one introduces the spin network states $\psi_{s}$, which minimizes the loop basis. Adopting these appropriate states, we can construct the area operator and find its eigenvalues. The area of the surface $\varSigma$ is rewritten in terms of the triad density

\begin{equation}
A_{\Sigma}=\int_{\varSigma}dx^{1}dx^{2}\sqrt{\det\left(q^{\left(2\right)}\right)}=\int_{\varSigma}dx^{1}dx^{2}\sqrt{\tilde{E}_{i}^{3}\tilde{E}^{3i}}.
\end{equation}

\noindent To express this area as the area operator, we should express the triad $\tilde{E}_{i}^{3}$ as the operator $\hat{\tilde{E}}_{i}^{3}$
by Eq. (\ref{eq:triad operator}) at first. Then the area operator takes a form as

\begin{equation}
\hat{A}_{\Sigma}=\int_{\varSigma}dx^{1}dx^{2}\sqrt{\left[\hat{\tilde{E}}_{i}^{3}\right]_{f}\left[\hat{\tilde{E}}_{i}^{3}\right]_{f}}.
\end{equation}

\noindent In the above equation, $f$ represents smearing functions. Impose this operator on the spin network states $\psi_{s}$. Thus we get

\begin{equation}
\hat{A}_{\Sigma}\psi_{s}=8\pi\ell_{p}^{2}\gamma\underset{I}{\sum}\sqrt{j_{I}\left(j_{I}+1\right)}\psi_{s},
\end{equation}

\noindent where $I$ denotes the edge of the spin networks and $j_{I}$ is a positive half integer. In string theory, we quantize the world sheet in the Ricci flat target space. The result of the closed string includes the massless spin-2 graviton naturally. The interaction of these massless closed strings curves the spacetime background. However, in LQG, we do not quantize a field in any other backgrounds, but use a quanta of the space to construct the background itself. Therefore, the area of the space is discrete and there is the minimal area which can be seen as a ground state of the quantum system. The minimal area is

\begin{equation}
A_{\mathrm{min}}=4\pi\sqrt{3}\gamma\ell_{p}^{2}.
\end{equation}

\noindent It is an order of $\ell_{p}^{2}$, which implies that there is a minimal length $\sim\ell_{p}$ . The review paper for loop quantum gravity and loop quantum cosmology can be seen in Refs. \cite{Bojowald:2008zzb,Rovelli:2010bf,Dona:2010hm,Banerjee:2011qu,Bilson-Thompson:2014hoa}.

The minimal observable length derived from GUP has been discussed in the introduction. We do not need to review it here.

\section{Review the tunneling method}

\subsection{The null geodesic method}
In the null geodesic method, the imaginary part of the action is derived by the particle's geodesic \cite{KW1,KW2,PW}. The energy conservation is considered, thus the background spacetime is varied with the particle's emission. We take the tunneling radiation in the Schwarzschild black hole for instance to review this method. The Schwarzschild metric is given by

\begin{eqnarray}
ds^2 = -f(r)dt^2 + \frac{1}{g(r)}dr^2 + r^2 (d\theta^2 + sin^2\theta d\phi^2),
\label{eq2.1.1}
\end{eqnarray}

\noindent with $ f\left(r\right)=g\left(r\right)=1-\frac{2M}{r}, $ $ M $ is the black hole's mass. The event horizon is located at $ r_E = 2M $. To avoid the singularity at the horizon, we perform the Painleve coordinate transformation

\begin{eqnarray}
dt\rightarrow dt - \sqrt{\frac{1-g(r)}{f(r)g(r)}} dr ,
\label{eq2.1}
\end{eqnarray}

\noindent on the metric (\ref{eq2.1.1}) and get

\begin{eqnarray}
ds^2 = -\left(1-\frac{2M}{r}\right)dt^2 + 2\sqrt{\frac{2M}{r}}dtdr + dr^2 + r^2 (d\theta^2 + sin^2\theta d\phi^2).
\label{eq2.1.2}
\end{eqnarray}

\noindent The above metric describes a stationary (not a static) spacetime and the singularity at the horizon is eliminated. We assume an s-wave outgoing positive energy particle. The radial null geodesics of the particle are defined by $ds^2=0$ from the metric (\ref{eq2.1.2}), which are

\begin{eqnarray}
\dot r = \frac{dr}{dt}= \pm 1- \sqrt{\frac{2M}{r}},
\label{eq2.1.3}
\end{eqnarray}

\noindent where $+ (-)$ correspond to the outgoing (ingoing) geodesics. We fix the total energy of a spacetime and allow that of the black hole to fluctuate. When a particle with energy $\omega$ is emitted, the energy of the black hole's system should be reduced. Then the mass $M$ should be replaced by $M- \omega$ in Eqs. (\ref{eq2.1.2}) and (\ref{eq2.1.3}). Thus the geodesics become

\begin{eqnarray}
\dot r  = \pm 1- \sqrt{\frac{2(M - \omega)}{r}}.
\label{eq2.1.4}
\end{eqnarray}

\noindent The tunneling rate is determined by the imaginary part of the emitted particle's action. The imaginary part of the action can be expressed as

\begin{eqnarray}
ImS = Im \int _{r_{in}}^{r_{out}} p_r dr= Im \int _{r_{in}}^{r_{out}} \int_{0}^{p_r}  dp^{\prime} _r dr,
\label{eq2.1.5}
\end{eqnarray}

\noindent where  $r_{in}$ and $r_{out}$ are the initial and final radii of the black hole, respectively. To solve the above equation, we introduce the Hamilton's equation $\dot r = \frac{dH}{dp_r}\mid _r$, and $H=M-\omega^{\prime}$. Then the above equation becomes

\begin{eqnarray}
ImS &=& Im \int _{r_{in}}^{r_{out}} \int_{M}^{M-\omega} \frac{dr}{\dot r} d(M- \omega^{\prime}) \nonumber\\
&=& Im \int _{r_{in}}^{r_{out}} \int_{M}^{M-\omega} \frac{dr}{1-\sqrt{\frac{2(M-\omega^{\prime})}{r}}} d(M- \omega^{\prime})\nonumber\\
&=& 4\pi M\omega \left(1-\frac{\omega}{2M}\right).
\label{eq2.1.6}
\end{eqnarray}

\noindent Therefore, the tunneling rate of the massless particle across the horizon is

\begin{eqnarray}
\Gamma \sim exp(-2ImS)=exp\left[-8\pi M \omega \left(1-\frac{\omega}{2M}\right)\right]=exp(\Delta S_{B-H}),
\label{eq2.1.7}
\end{eqnarray}

\noindent where $\Delta S_{B-H}$ is the change of the Bekenstein-Hawking entropy. This equation is the Boltzmann factor with the Hawking temperature at the horizon taking

\begin{eqnarray}
T=\frac{1}{8\pi M(1-\frac{\omega}{2M})} = (1+\frac{\omega}{2M})T_0.
\label{eq2.1.8}
\end{eqnarray}

\noindent In the above equation, $T_0 = \frac{1}{8\pi M}$ is the original Hawking temperature derived without consideration of the varied background spacetime. Clearly, the true temperature is higher than the original one. The original Hawking formula predicts the complete evaporation of the black hole. Therefore, we can conclude that the varied spacetime accelerates the evaporation.

The above discussion is focused on the massless particle. Zhang and Zhao extended this work the case of massive particles \cite{ZZ1,ZZ2,JWC}. For a massive particle, its equation of motion is different from that of the massless particle. According to de Broglie��s hypothesis and the definition of the phase velocity $v_p$ and the group $v_g$ velocity, we have the relationship

\begin{eqnarray}
v_p =\frac{dr}{dt} =\dot r, \quad v_g = \frac{dr_c}{dt}, \quad v_p = \frac{1}{2}v_g,
\label{eq2.1.9}
\end{eqnarray}

\noindent where $r_c$ is the radial position of the particle. We still carry out the discussion in the Painleve metric (\ref{eq2.1.2}). Using Landau's theory of the coordinate clock synchronization, we get the coordinate time difference of these two events

\begin{eqnarray}
dt= -\frac{g_{01}}{g_{00}}dr_c, \quad (d\theta = d\varphi = 0).
\label{eq2.1.10}
\end{eqnarray}

\noindent Then the group velocity is gotten as

\begin{eqnarray}
v_g = \frac{dr_c}{dt} =-\frac{g_{00}}{g_{01}},
\label{eq2.1.11}
\end{eqnarray}

\noindent and the phase velocity is

\begin{eqnarray}
v_p=\dot r =\frac{1}{2}v_g =-\frac{g_{00}}{g_{01}} = \frac{1-\frac{2M}{r}}{\sqrt{\frac{2M}{r}}}.
\label{eq2.1.12}
\end{eqnarray}

\noindent Inserting the above equation into the imaginary part (\ref{eq2.1.5}) and adopting the Hamilton's equation $\dot r = \frac{dH}{dp_r}\mid _r$, we can get the same imaginary part as Eq. (\ref{eq2.1.6}). Therefore, the tunneling rate and the Hawking temperature are recovered by the massive particle across the horizon.

\subsection{The Hamilton-Jacobi method}

The embryonic form of the Hamilton-Jacobi method is the complex integration method \cite{SP,SPS,ANVZ}. In this method, we don't need to assume that a particle moves radially. The action of the particle is derived by the Hamilton-Jacobi equation
\begin{eqnarray}
g^{\mu\nu} \partial_{\mu} I\partial_{\nu} I + m^2 = 0,
\label{eq2.2.1}
\end{eqnarray}

\noindent where $m$ and $I$ are the mass and the action of the particle. Here we adopt the Schwarzschild metric (\ref{eq2.1.1}) to review this method. Inserting the inverse metric of the metric (\ref{eq2.1.1}) into the Hamilton-Jacobi equation yields

\begin{eqnarray}
-\frac{1}{f} (\partial_{t} I)^2 +g (\partial_{r} I)^2+ \frac{1}{r^2} (\partial_{\theta} I)^2+ \frac{1}{r^2sin^2\theta} (\partial_{\phi} I)^2 + m^2 = 0.
\label{eq2.2.2}
\end{eqnarray}

\noindent It is difficult to solve the above equation directly. Taking into account the properties of the metric, we carry out variables of separation as follows

\begin{eqnarray}
I = -\omega t + W(r) + J(\theta, \phi),
\label{eq2.2.3}
\end{eqnarray}

\noindent where $\omega$ is the energy of the emitted particle. Substituting the separated variables into Eq. (\ref{eq2.2.2}) and solving it yield

\begin{eqnarray}
W_{\pm} &=& \pm \int {\frac{dr}{\sqrt{gf}} \sqrt{\omega^2 -f\left[\frac{1}{r^2} (\partial_{\theta} J)^2+ \frac{1}{r^2sin^2\theta} (\partial_{\phi} J)^2 + m^2\right]}}\nonumber\\
&=&  \pm i \pi 2M\omega,
\label{eq2.2.4}
\end{eqnarray}

\noindent where $+(-)$ correspond to the outgoing (ingoing) solutions \cite{Mitra}. Thus the tunneling rate is

\begin{eqnarray}
\Gamma & = & \frac{P_{(emission)}}{P_{(absorption)}} = \frac{\exp(-2Im I_+ )}{\exp(-2Im I_-)}= \frac{\exp[-2Im W_+ - 2Im J]}{\exp[-2Im W_- - 2Im J]}\nonumber\\
& = & \exp\left(-8\pi M\omega\right).
\label{eq2.2.5}
\end{eqnarray}

\noindent In the above equation, the solution of $J$ is the complex function, but the contribution was canceled in the calculation. Therefore, the tunneling rate is only related to the radial action of the particle. Thus the Hawking temperature of the black hole is gotten as

\begin{eqnarray}
T=\frac{1}{8\pi M}.
\label{eq2.2.6}
\end{eqnarray}

\noindent In the discussion, since we neglected the varied background spacetime of the black hole, the standard Hawking temperature, which implies the complete evaporation of the black hole, was recovered. When the varied spacetime are taken into account, the corrected Hawking temperature (\ref{eq2.1.8}) will be gotten \cite{CY,MPK11}.

\subsection{Fermions' tunneling radiation}

In this subsection, we adopt the Hamilton-Jacobi method to review the tunneling radiation of fermions across the horizon of the Schwarzschild spacetime \cite{KM2,KM3}. The equation of motion of a spin-1/2 particle satisfies $i\gamma^{\mu}\left(\partial_{\mu}+\Omega_{\mu} \right)\psi+\frac{m}{\hbar}\psi=0$, which is used in the signature metric $(+,-,-,-)$. When we use the signature $(-,+,+,+)$, the equation of motion should be \footnote{Please see section 36: Lagrangians for spinor fields, ``Quantum Field Theory'', M. Srednicki. It was also discussed in Ref. \cite{liuren2}.}

\begin{eqnarray}
\gamma^{\mu}\left(\partial_{\mu}+\Omega_{\mu} \right)\psi+\frac{m}{\hbar}\psi=0,
\label{eq2.3.1}
\end{eqnarray}

\noindent where $m$ is the mass of the fermion, $\Omega _\mu \equiv\frac{i}{2}\omega _\mu\, ^{a b} \Sigma_{ab}$, $\omega _\mu\, ^{ab}$ is the spin connection and is defined by the ordinary connection $\omega_\mu\,^a\,_b=e_\nu\,^a e^\lambda\,_b \Gamma^\nu_{\mu\lambda} -e^\lambda\,_b\partial_\mu e_\lambda\,^a$, $e^a_{\mu}$ is a tetrad. The Greek indices are raised and lowered by the curved spacetime metric $g_{\mu\nu}$ and the Latin indices are governed by the flat metric $\eta_{ab}$. The construction of the tetrad relies on the definitions

\begin{equation}
g_{\mu\nu}= e_\mu\,^a e_\nu\,^b \eta_{ab},\hspace{5mm} \eta_{ab}=
g_{\mu\nu} e^\mu\,_a e^\nu\,_b, \hspace{5mm} e^\mu\,_a e_\nu\,^a=
\delta^\mu_\nu, \hspace{5mm} e^\mu\,_a e_\mu\,^b = \delta_a^b.
\label{eq2.3.2}
\end{equation}

\noindent $\Sigma_{ab}$ is the Lorentz spinor generator defined by

\begin{equation}
\Sigma_{ab}= \frac{i}{4}\left[ {\gamma ^a ,\gamma^b} \right],
\hspace{5mm} \{\gamma ^a ,\gamma^b\}= 2\eta^{ab}. \label{eq2.3.3}
\end{equation}

\noindent Then it is easily to construct the $\gamma^\mu$ in curved spacetime, which is

\begin{equation}
\gamma^\mu = e^\mu\,_a \gamma^a, \hspace{7mm} \left\{ {\gamma ^\mu
,\gamma ^\nu } \right\} = 2g^{\mu \nu }. \label{eq2.3.4}
\end{equation}

To deal with Hawking radiation of fermions, one should first choose a tetrad. It is straightforward to guess the tetrad from the metric (\ref{eq2.1.1}). It is

\begin{equation}
e_{\mu}^{\quad a}=\mathrm{diag}\left(\sqrt{f},1/\sqrt{g},r,r\sin\theta\right).
\label{eq2.3.4}
\end{equation}

\noindent Then the gamma matrices are given by

\begin{eqnarray}
\gamma ^t &=& \frac{1}{\sqrt
{f} }\left( {{\begin{array}{*{20}c}
 i \hfill & 0 \hfill \\
 0 \hfill & { - i} \hfill \\
\end{array} }} \right),
\quad \gamma ^\theta = \frac{1}{r}\left(
{{\begin{array}{*{20}c}
 0 \hfill & {\sigma ^1} \hfill \\
 {\sigma ^1} \hfill & 0 \hfill \\
\end{array} }} \right), \nonumber\\
\gamma ^r &=& \sqrt {g} \left(
{{\begin{array}{*{20}c}
 0 \hfill & {\sigma ^3} \hfill \\
 {\sigma ^3} \hfill & 0 \hfill \\
\end{array} }} \right),
\quad \gamma ^\phi = \frac{1}{r\sin \theta }\left(
{{\begin{array}{*{20}c}
 0 \hfill & {\sigma ^2} \hfill \\
 {\sigma ^2} \hfill & 0 \hfill \\
\end{array} }} \right).
\label{eq2.3.6}
\end{eqnarray}

\noindent Here, $\sigma ^i$'s are the Pauli Sigma matrices given by $\sigma^{1}=\left(\begin{array}{cc} 0 & 1\\1 & 0\end{array}\right) $, $\sigma ^{2}=\left(\begin{array}{cc} 0 & -i\\ i & 0 \end{array}\right)$, $\sigma ^{3}=\left(\begin{array}{cc} 1 & 0\\ 0 & -1 \end{array}\right)$. The spin-1/2 fermion has two spin states, namely, spin up state and spin down state. Their wave functions are expressed as

\begin{eqnarray}
\Psi _ \uparrow &=& \left( {{\begin{array}{*{20}c}
 A \left({t,r,\theta ,\phi } \right) \hfill \\
 0 \hfill \\
 B\left({t,r,\theta ,\phi } \right) \hfill \\
 0 \hfill \\
\end{array} }} \right)\exp \left( {\frac{i}{\hbar }I_ \uparrow \left(
{t,r,\theta ,\phi } \right)} \right), \nonumber\\
 \Psi _ \downarrow &=& \left( {{\begin{array}{*{20}c}
 0 \hfill \\
 C\left({t,r,\theta ,\phi } \right) \hfill \\
 0 \hfill \\
 D\left({t,r,\theta ,\phi } \right) \hfill \\
\end{array} }} \right)\exp \left( {\frac{i}{\hbar }I_ \downarrow \left(
{t,r,\theta ,\phi } \right)} \right),\label{eq2.3.8}
\end{eqnarray}

\noindent where $\uparrow$ denotes the case of the spin up state, $\downarrow$ is for spin down state case, and $I$ is the action of the fermion. In this paper, we only discuss the tunneling radiation of the fermion with spin up state. The process of the spin down state is parallel. Inserting the ansatz Eq. (\ref{eq2.3.8}) for the spin up state into the Dirac equation, dividing the exponential term and multiplying by $\hbar$, we get four equations

\begin{eqnarray}
- \left( {\frac{iA}{\sqrt {f}} {\partial _t I_ \uparrow  }  + B\sqrt {g} \partial _r I_ \uparrow } \right) + imA &=& 0, \label{eq2.3.9}
\\
\left( {\frac{iB}{\sqrt {f} } {\partial _t I_ \uparrow }  - A\sqrt {g} \partial _r I_ \uparrow } \right) + imB &=& 0, \label{eq2.3.10}
\\
- B\left( {\frac{1}{r}\partial _\theta I_ \uparrow + \frac{i}{r\sin \theta }\partial _\phi I_ \uparrow } \right) &=& 0,\label{eq2.3.11}\\
 - A\left( {\frac{1}{r}\partial _\theta I_ \uparrow + \frac{i}{r\sin \theta }\partial _\phi I_ \uparrow } \right) &=& 0. \label{eq2.3.12}
\end{eqnarray}

Our aim is to get the solution of the action. However, it is difficult to solve the action from the above equations. Considering the symmetries of the black hole spacetime, we carry out separation of variables for the action as

\begin{eqnarray}
I_\uparrow= - \omega t + W\left( r \right) + J \left( {\theta,\phi } \right), \label{eq2.3.13}
\end{eqnarray}

\noindent where $\omega $ is the energy of the emitted fermion. Eqs. (\ref{eq2.3.11}) and (\ref{eq2.3.12}) are the same equation and independent on $A$ and $B$. Inserting Eq. (\ref{eq2.3.13}) into Eqs. (\ref{eq2.3.11}) and (\ref{eq2.3.12}) yields a complex function solution $J$ (other than a constant solution). Although $J$ is solved as a complex function, and then would provide a contribution to the imaginary part of the action, its total contributions to the tunneling rate will be canceled out. Therefore, we neglect it here. Then substituting Eq. (\ref{eq2.3.13}) into Eqs. (\ref{eq2.3.9}) and (\ref{eq2.3.10}), we get the radial function

\begin{eqnarray}
\left( {\frac{iA}{\sqrt {f}}\omega  - B\sqrt {g} \partial _r W} \right) + imA &=& 0,
\\
\left( {\frac{iB}{\sqrt {f} } \omega +A\sqrt {g} \partial _r W} \right) - imB &=& 0.\label{eq2.3.14}
\end{eqnarray}

\noindent  At the outer horizon, $f(r)=g(r) = 0$. The solution of the radial function $W\left( r \right)$ can be gotten from the above equations and is

\begin{eqnarray}
 W_\pm  & =& \pm \int {\frac{\sqrt {\omega ^2 - m^2f}
}{\sqrt{gf}}dr} \nonumber\\
& =& \pm \int {\frac{\sqrt {\omega ^2 - m^2\left(1-2M/r\right)}
}{1-2M/r}dr} \nonumber\\
&=&\pm i\pi 2M \omega, \label{eq2.3.15}
\end{eqnarray}

\noindent where $+ $/$-$ correspond to the outgoing/ingoing solutions \cite{Mitra}. So the tunneling rate of the fermion across the horizon is

\begin{eqnarray}
 \Gamma &=& \frac{P_{\left( {emission} \right)} }{P_{\left( {absorption}
\right)} } = \frac{\exp ( - 2\textrm{Im}I_ {\uparrow+ })}{\exp
( - 2\textrm{Im}I_{\uparrow - })}= \frac{\exp ( - 2\textrm{Im}W_ + )}{\exp ( - 2\textrm{Im}W_ - )}\nonumber\\
& =& \exp \left( { - 8\pi M \omega} \right), \label{eq2.3.16}
\end{eqnarray}

\noindent which means the Hawking temperature of the black hole taking

\begin{equation}
T = \frac{1}{8\pi M }. \label{eq2.3.17}
\end{equation}

\noindent Therefore, the Hawking temperature can be recovered by the fermion tunneling across the horizon of the black hole.

In this section, we have reviewed the tunneling method. Due to the varied background spacetime, the null geodesic method yielded the leading correction to the Hawking temperature. In the Hamilton-Jacobi method, we considered the fixed spacetime, thus the standard Hawking temperature were recovered. All of these results predicted the complete evaporation of the black hole. In next section, we adopt the new definition of the geodesic equation to discuss the tunneling radiation.

\section{New derivation of the geodesic equation}

The geodesic equation plays an important role in the null geodesic method to research on Hawking radiation. However, the derivation of the geodesic equation of the massive particle, which is different from that of the massless case, is unnatural. The geodesic equation of the massless particle is defined by $ds^2=0$, while that of the massive particle is defined by the relation between the group velocity and the phase velocity. This derivation is not in consistence with the first principle--the variation principle. General relativity tells us that both of the geodesic equations of massless and massive particles can be derived directly by the variation principle on the Lagrangian action. In this section, we derive the geodesic equations of the massless and massive particles from the variation principle, and then use these equations to derive the tunneling radiation of a black string. This method was first developed by Wu et al.. \cite{WJW,JCW}.

The metric of the black string was obtained from Einstein field equations with a negative cosmological constant. It describes a cylindrically symmetric spacetime and is given by \cite{JPS}

\begin{eqnarray}
ds^2 = -\Delta dt^2 + \Delta^{-1}dr^2 + r^2 d\theta^2 +\alpha ^2 r^2 dz^2,
\label{eq3.1}
\end{eqnarray}

\noindent where $\Delta = \alpha^2 r^2 -\frac{M}{r}$, $0 \leq \theta \leq 2\pi$, $0 \leq r < \infty$, $-\infty < z < \infty$, $\alpha^2= -\Lambda /3$, $\Lambda$ is the cosmological constant, and $M$ is related to the ADM mass density. The event horizon locates at $r_+ = \left(\frac{M}{\alpha^2}\right)^{1/3}$. The entropy of the black string is $S= \frac{1}{2} \pi \alpha r_+^2$. In the following, we use the Lagrangian analysis on the action to get the geodesic equation of the emitted particle in the Painleve coordinate system. Therefore, the above metric should be rewritten as

\begin{eqnarray}
ds^2 = -\Delta dt^2 +2\sqrt{1-\Delta} dtdr +dr^2 + r^2 d\theta^2 +\alpha ^2 r^2 dz^2,
\label{eq3.2}
\end{eqnarray}

\noindent which was gotten by the Painleve coordinate transformation $dt\rightarrow dt -  \sqrt{1-\Delta}\Delta^{-1}dr $. When a massive (massless) particle tunnels across the horizon of the black string (\ref{eq3.2}), its Lagrangian quantity is

\begin{eqnarray}
L =\frac{1}{2}\left[-\Delta \dot t^2- 2 \sqrt{1-\Delta}\dot t^2\dot r^2 + \dot r^2+ r^2\dot {\theta}^2 + \alpha ^2 r^2\dot {z}^2\right].
\label{eq3.3}
\end{eqnarray}

\noindent In the above equation, the dot expresses the differentiation with respect to the affine parameter $\tau$. Then the canonical momenta of the particle are gotten as follows

\begin{eqnarray}
p_t &=& \frac{\partial L}{\partial \dot t}= -\Delta \dot t- \sqrt{1-\Delta}\dot r, \nonumber\\
p_r &=&  \frac{\partial L}{\partial \dot r}= \dot r- \sqrt{1-\Delta}\dot t, \nonumber\\
p_{\theta}&=&  \frac{\partial L}{\partial \dot \theta}= r^2 \dot \theta, \nonumber\\
p_z &=& \frac{\partial L}{\partial \dot z}= \alpha^2 r^2\dot z.
\label{eq3.4}
\end{eqnarray}

\noindent After the Legendre transformation, we get the Hamiltonian quantity

\begin{eqnarray}
H =\frac{1}{2}\left[-\Delta \dot t^2- 2 \sqrt{1-\Delta}\dot t^2\dot r^2 + \dot r^2+ r^2\dot {\theta}^2 + \alpha ^2 r^2\dot {z}^2\right],
\label{eq3.5}
\end{eqnarray}

\noindent which is constrained to a constant $-k/2$. $k=0$ corresponds to the case for the geodesic equation of the massless particle, while $k=1$ denotes that of the massive particle. In the Lagrangian (\ref{eq3.3}), $t$ and $z$ are the cyclic coordinates. Thus we get $\frac{dp_t}{d \tau} =\frac{dL}{d t} = 0$ and $\frac{dp_z}{d \tau} =\frac{dL}{d z} =0$, which means that $p_t$ and $p_z$ should be constants. Let $p_t = E$ and $p_z = J$, where $E$ and $J$ are constants of integration. To solve $\dot t$, $\dot r$, without loss of generality, we let $\theta$ be fixed at a certain angular. Combining Eqs. (\ref{eq3.4}) and (\ref{eq3.5}), we get

\begin{eqnarray}
\dot r &=& \pm \sqrt{E^2 - \Delta \left(k+\frac{J^2}{\alpha^2r^2}\right)} , \nonumber\\
\dot t &=&  \frac{1}{\Delta} \left[-E \pm \sqrt{1-\Delta}\sqrt{E^2 - \Delta \left(k+\frac{J^2}{\alpha^2r^2}\right)} \right].
\label{eq3.6}
\end{eqnarray}

\noindent Thus the geodesic equation of the massive (massless) particle across the horizon can be easily gotten. It is

\begin{eqnarray}
\bar{r} \equiv \frac{dr}{dt}=\frac{\dot r}{\dot t } =\frac{\Delta \sqrt{E^2 - \Delta \left(k+\frac{J^2}{\alpha^2r^2}\right)}}{\pm E - \sqrt{1-\Delta}\sqrt{E^2 - \Delta \left(k+\frac{J^2}{\alpha^2r^2}\right)} },
\label{eq3.7}
\end{eqnarray}

\noindent where $+ (-)$ denote the (outgoing) ingoing geodesics. Let the particle cross the horizon with the s-wave way. When $k = 0$, we get $\bar{r} = 1+ \sqrt{1-\Delta}$, which describes the geodesic equation of the massless particle and is full in consistence with that derived by $ds^2=0$. Following Parikh and Wilczek's work, we can easily get the tunneling rate of the massless particle across the horizon of the black string. The result is consistent with the results gotten in Parikh and Wilczek's work \cite{PW}. When $k=1$, the geodesic equation of the massive particle is $\bar{r} = \Delta \left[-\sqrt{1-\Delta}\pm\frac{ E}{ {\sqrt{E^2 - \Delta }}}\right]^{-1}$. Now we use the geodesic equation (\ref{eq3.7}) to calculate the tunneling rate of the massive (massless) particle across the horizon. When a particle with energy $\omega$ is emitted, the energy of the system of the black string should be reduced as $M- \omega$. Taking into account the self-gravity of the particle, we should replace $M$ with $M- \omega$ in the metric (\ref{eq3.2}) and in the geodesic equation (\ref{eq3.7}). Thus we rewrite Eq. (\ref{eq3.7}) as

\begin{eqnarray}
\bar{r} = \tilde \Delta \left[-\sqrt{1-\tilde\Delta}\pm\frac{ \tilde E}{ {\sqrt{ \tilde E^2 - \tilde\Delta k}}}\right]^{-1},
\label{eq3.8}
\end{eqnarray}

\noindent where $\tilde \Delta = \alpha^2 r^2 -\frac{M-\omega}{r}$, $\tilde E$ is the integration constant after the particle emission with consideration of the self-gravitation. The tunneling rate is determined by the imaginary part of the action of the particle. We follow the process in the section 2 and adopt the expression of Eq. (\ref{eq2.1.5}) and the Hamilton's equation $\bar r = \frac{dH}{dp_r}\mid _r= \frac{d(M-\omega^{\prime})}{dp_r}\mid _r$. The imaginary part is gotten as

\begin{eqnarray}
ImS &=& Im \int _{r_{in}}^{r_{out}} \int_{0}^{p_r}  dp^{\prime} _r dr= Im \int _{r_{in}}^{r_{out}} \int_{M}^{M-\omega} \frac{dr}{\bar r} d(M- \omega^{\prime}) \nonumber\\
&=& Im \int _{r_{in}}^{r_{out}} \int_{M}^{M-\omega} \frac{-\sqrt{1-\tilde\Delta}\pm\frac{ \tilde E}{ {\sqrt{ \tilde E^2 - \tilde\Delta k}}}}{\tilde \Delta } drd(M- \omega^{\prime})\nonumber\\
&=& -\frac{1}{4}\pi \alpha\left(r_{out}^2-r_{in}^2\right).
\label{eq3.9}
\end{eqnarray}

\noindent Clearly, the imaginary part was derived by the emissions of the massless and massive particles in the same equation. The imaginary part is irrelevant to the mass of the emitted particle. Therefore, the tunneling rate of the particle across the horizon is

\begin{eqnarray}
\Gamma \sim exp(-2ImS)=exp\left[ -\frac{1}{2}\pi \alpha\left(r_{out}^2-r_{in}^2\right)\right]=exp(\Delta S_{B-H}),
\label{eq3.10}
\end{eqnarray}

\noindent which is related to the change of the Bekenstein-Hawking entropy, $\Delta S_{B-H}$. This result is precisely in consistence with that derived in Refs. \cite{PW,ZZ1,ZZ2,JWC} and leads to the correction to the pure thermal spectrum.

\section{Tunneling radiation with corrections of quantum gravity}

In fact, the standard Hawking temperature can be directly obtained by the Heisenberg uncertainty principle (HUP) \cite{ACS,OR}. When a particle with positive energy $E$ is emitted to infinity, a particle with negative energy $-E$ is absorbed by the black hole. We can estimate the characteristic energy $E$ of the emitted particle across the horizon of the Schwarzschild black hole by HUP. In the vicinity of the black hole's surface, there is an position uncertainty of the order of the Schwarzschild radius, which is $\Delta x \simeq r_E = \frac{2GM}{c^2}$. Using the uncertainty principle

\begin{eqnarray}
\Delta x \Delta p \sim \hbar,
\label{1.1}
\end{eqnarray}

\noindent we get the momentum uncertainty as $\Delta p \simeq \frac{\hbar}{r_E} = \frac{\hbar c^2}{2GM }$. Thus the uncertainty in the energy of the emitted particle is

\begin{eqnarray}
\Delta E = c \Delta p = \frac{\hbar c^3}{2GM}.
\label{1.2}
\end{eqnarray}

\noindent We identify this energy as the characteristic energy of the emitted particle. Using a calibration factor $4\pi$ and setting $k_B = 1$, we can easily get the temperature as the following form

\begin{eqnarray}
T = \frac{\Delta E}{4\pi} = \frac{\hbar c^3}{8\pi GM},
\label{1.3}
\end{eqnarray}

\noindent which is the thermal spectrum of a black body radiation and shows the complete evaporation of the black hole. This result can be seen as a direct consequence of HUP. All the above results show the complete evaporation of the black holes.

However, various theories of quantum gravity predict the minimal observable length. When effects of quantum gravity are taken into account, there should be a minimal observable length for a black hole. It implies that the black hole can not evaporate completely. Introducing effects of quantum gravity by the modified de Broglie relation, the tunneling radiation of scalar particles in the varied background spacetime has been discussed in the work \cite{NS}. The corrected tunneling temperature was gotten. Through the modified Dirac equation, fermions' tunneling radiation in the various spacetimes has been discussed in our work \cite{DWY1,DWY2,DJWY,CHEN}. It was showed that the corrected temperatures are related to the quantum numbers of the emitted fermions. In this section, we incorporate effects of quantum gravity into black hole physics and discuss the tunneling radiation of scalar particles and fermions. The remnants are observed.

\subsection{Tunneling radiation of scalar particles}
\subsubsection{Quantum gravity corrections by GUP}

Doubly Special Relativity shows that the existence of a minimal measurable length would restrict a maximal momentum uncertainty of a particle \cite{GAC1,GAC2,JKG,IK}. This implies the existence of the maximal momentum since the minimal measurable length can be found at the Planck scale. Both of the minimal length and maximal momentum can be realized in the following DSR-GUP\cite{ADV}

\begin{eqnarray}
\Delta x \Delta p \geq \frac{\hbar}{2} \left(1 - \alpha_1 \Delta p+ \alpha_2 \Delta p^2\right),
\label{1.4}
\end{eqnarray}

\noindent where $\alpha_1 = 2\alpha_0 l_p /{\hbar}$,  $\alpha_2= 4\alpha_0^2 l_p^2/{\hbar^2}$, and $\alpha_0$ is a dimensionless parameter. The above GUP is derived by the modified fundamental commutation relations

\begin{eqnarray}
[x_i,p_j]= i\hbar\left\{\delta_{ij}-\alpha_0 l_p\left( p\delta_{ij}+p_ip_j/p\right)+\alpha_0^2 l_p^2\left(p^2\delta_{ij}+3p_ip_j\right)\right\},
\label{1.5}
\end{eqnarray}

\noindent which predicts a minimum measurable length and a maximum measurable momentum as $\Delta x \geq (\Delta x)_{min} \approx \alpha_0 l_p$ and $\Delta p \leq (\Delta p)_{max} \approx M_pc/{\alpha_0}$, respectively. $x_i$ and $p_i$ are operators of position and momentum defined by \cite{ADV}

\begin{eqnarray}
x_i &=& x_{0i}, \nonumber\\
p_i &=& p_{0i} (1 - \alpha_1 p_0 +2\alpha_1^2 p_0^2).
\label{1.6}
\end{eqnarray}

\noindent In the above equation, $x_{0i}$, $p_{0j}$ satisfy the canonical commutation relations. When $\alpha_1 =0$ and $\alpha_2 = \beta$, the above DSR-GUP reduces to the ordinary GUP expressed as Eq. (\ref{eq1.1}). Uing the DSR-GUP, the semi-classical tunneling radiation was reprocessed in the work \cite{NS}. When the DSR-GUP is taken into account, the usual de Broglie relation should be modified as

\begin{eqnarray}
\lambda\simeq \frac{1}{p}\left(1- \alpha_1  p+ \alpha_2  p^2\right),
\label{1.7}
\end{eqnarray}

\noindent or equivalently

\begin{eqnarray}
\omega \simeq E\left(1- \alpha_1  E+ \alpha_2 E^2\right).
\label{1.8}
\end{eqnarray}

\noindent On the other hand, from the DSR-GUP expression, we modify the commutation relation between the radial coordinate and the conjugate momentum as $[r,p_r]= i \hbar \left(1- \alpha_1  p+ \alpha_2 p^2\right)$. When $\alpha_2=\alpha_1^2 = \alpha_0^2 l_p^2/{\hbar^2}$, the above relation is reduced into that gotten in the reference \cite{NS}. In the classical limit, the following poisson bracket

\begin{eqnarray}
\{r,p_r\}= 1- \alpha_1  p+ \alpha_2  p^2,
\label{1.9}
\end{eqnarray}

\noindent is adopted to replace by the commutation relation. To get the imaginary part of the action, the following deformed Hamiltonian equation is adopted
\begin{eqnarray}
\dot r=\{r,H\}= \{r,p_r\}\frac{dH}{dp_r}\mid _r,
\label{1.10}
\end{eqnarray}

\noindent Following the work \cite{NS}, we let $p_r=\omega^{\prime}$ and $p_r^2={\omega^{\prime}}^2$. To eliminate the momentum in the favor of the energy in integral Eq. (\ref{eq2.1.5}), we insert Eqs. (\ref{1.10}) and (\ref{eq2.1.4}) into Eq. (\ref{eq2.1.5}) and get

\begin{eqnarray}
ImS &=& Im \int _{r_{in}}^{r_{out}} \int_{M}^{M-\omega} \frac{1- \alpha_1 \omega+ \alpha_2  \omega^2}{\dot r} drdH^\prime \nonumber\\
&=& Im \int _{r_{in}}^{r_{out}} \int_{M}^{M-\omega} \frac{1- \alpha_1 \omega+ \alpha_2 \omega^2 }{1-\sqrt{\frac{2(M-\omega^{\prime})}{r}}}drd(M- \omega^{\prime})\nonumber\\
&=&  4\pi M\omega \left[1- \frac{\omega}{2M} -\alpha_1\omega\left(\frac{1}{2}-\frac{\omega }{3M}\right)+\alpha_2 \omega^2\left(\frac{1}{3}-\frac{\omega}{4M}\right)  \right].
\label{1.11}
\end{eqnarray}

Therefore, the tunneling rate of the particle across the horizon with effects of quantum gravity is gotten as

\begin{eqnarray}
\Gamma \sim exp\left\{-8\pi M\omega \left[1- \frac{\omega}{2M} -\alpha_1\omega\left(\frac{1}{2}-\frac{\omega }{3M}\right)+\alpha_2 \omega^2\left(\frac{1}{3}-\frac{\omega}{4M}\right)  \right]\right\}.
\label{1.12}
\end{eqnarray}

\noindent In the above equation, there are four terms in the exponential. The first term gives a thermal Boltzmannian spectrum. The other three terms are correction values. The second term is caused by the varied back ground spacetime, which was found in the work of Parikh and Wilczek. The third and fourth terms are caused by the effects of quantum gravity. When $\alpha_1=0$, the above result reflects the tunneling rate of the varied spacetime influenced by the quantum gravity correction with the minimal length. When $\alpha_1=\alpha_2=0$, the above result reduces to that of Eq. (\ref{eq2.1.6}).

\subsubsection{The generalized Klein-Gordon equation}

In this subsection, we introduce GUP effects into Klein-Gordon equation in curved spacetime by the modified operators of position and momentum \cite{CHENWU}.
The derivation of GUP relies on the modified fundamental commutation relations. There are two representations to express the commutation relations. The first way is to modify the momentum operator in the position representation covariantly, which has been discussed in the introduction. The second way is to modify the position operator in the momentum representation. This will be given by Eq. (\ref{1.29}). In this section, we adopt the position representation.

Using Eq. (\ref{eq1.2}) and neglecting the higher order term of $\beta$, we get

\begin{eqnarray}
p^2 &=& p_i p^i = -\hbar^2 \left[ {1 - \beta \hbar^2 \left(
{\partial _j \partial ^j} \right)} \right]\partial _i \cdot \left[
{1 - \beta \hbar^2 \left( {\partial ^j\partial _j } \right)}
\right]\partial ^i\nonumber \\
&\simeq & - \hbar ^2\left[ {\partial _i \partial ^i - 2\beta \hbar
^2 \left( {\partial ^j\partial _j } \right)\left( {\partial
^i\partial _i } \right)} \right], \label{eq4.1.1}
\end{eqnarray}

\noindent To take into account effects of quantum gravity  \cite{WG}, we introduce the generalized frequency

\begin{eqnarray}
\tilde \omega = E( 1 - \beta E^2). \label{eq4.1.2}
\end{eqnarray}

\noindent The energy operator in the above equation is defined by $ E = i \hbar \partial _t $. Using the energy mass shell condition $ p^2 + m^2 = E^2 $, it is easy to derive the modified expression of energy \cite{WG,NS,NK,HBHRSS}

\begin{eqnarray}
\tilde E = E[ 1 - \beta (p^2 + m^2)]. \label{eq4.1.3}
\end{eqnarray}

\noindent In curved spacetime, the Klein-Gordon equation takes the form as

\begin{eqnarray}
g^{\mu\nu}p_{\mu}p_{\nu}\Psi=-m^{2}\Psi,
\label{KG}
\end{eqnarray}

\noindent To introduce the effects of quantum gravity, we rewrite the above equation as

\begin{eqnarray}
-\left(i\hbar\right)^{2}\partial^{t}\partial_{t}\Psi=\left[\left(i\hbar\right)^{2}\partial^{i}\partial_{i}+m^{2}\right]\Psi.
\label{eq:KG}
\end{eqnarray}

\noindent Substituting Eqs. (\ref{eq4.1.1}) and (\ref{eq4.1.3}) into Eq. (\ref{eq:KG}) and keeping the leading order of $\beta$, we get the generalized Klein-Gordon equation

\begin{eqnarray}
-\left(i\hbar\right)^{2}\partial^{t}\partial_{t}\Psi=\left[\left(i\hbar\right)^{2}\partial^{i}\partial_{i}+m^{2}\right]\left[1-2\beta\left(\left(i\hbar\right)^{2}
\partial^{i}\partial_{i}+m^{2}\right)\right]\Psi.\label{eq:gen KG}
\end{eqnarray}

\noindent The effect of an electromagnetic field was not considered in the above equation. If we consider this effect, the generalized Klein-Gordon equation (\ref{KG}) should be changed.

\subsubsection{Scalar particles' tunneling radiation in the Schwarzschild black hole}

In this subsection, we use the generalized Klein-Gordon equation to discuss the tunneling radiation of a scalar particle in the Schwarzschild black hole. The Schwarzschild metric is given by Eq. (\ref{eq2.1.1}). The equation of motion of a scalar particle satisfies the generalized Klein-Gordon equation (\ref{eq:gen KG}). To realize the quantum property of the scalar particle and utilize the WKB approximation, we assume that the wave function of the particle is

\noindent
\begin{equation}
\Psi=\exp\left[\frac{i}{\hbar}I\left(t,r,\theta,\phi\right)\right],\label{eq:SB wave}
\end{equation}

\noindent where $I$ is the action of the scalar particle. To use the WKB approximation, we substitute the wave function (\ref{eq:SB wave}) into the Klein-Gordon equation (\ref{eq:gen KG}) and combine the metric in Eq. (\ref{eq2.1.1}). When the higher orders of $\hbar$ and $\beta$ are neglected, we get

\begin{eqnarray}
\frac{1}{f}\left(\partial_{t}I\right)^{2} & = & \left[\left(g\left(\partial_{r}I\right)^{2}+\frac{1}{r^{2}}\left(\partial_{\theta}I\right)^{2}+\frac{1}{r^{2}\sin^{2}\theta}\left(\partial_
{\phi}I\right)^{2}\right)+m^{2}\right]\times\nonumber \\
 &  & \left(1-2\beta\left[g\left(\partial_{r}I\right)^{2}+\frac{1}{r^{2}}\left(\partial_{\theta}I\right)^{2}+\frac{1}{r^{2}\sin^{2}\theta}\left(\partial_
 {\phi}I\right)^{2}+m^{2}\right]\right). \label{eq:SB I}
\end{eqnarray}

\noindent It is difficult to solve the above equation. Considering the properties of the spacetime, we carry out separation of variables on the action

\noindent
\begin{equation}
I=-\omega t+W\left(r\right)+J\left(\theta,\phi\right),\label{eq:SB sep}
\end{equation}

\noindent where $\omega$ is the energy of the emitted particle. Following the previous work, we choose the s-wave \cite{PW}, which means that $\frac{1}{r^{2}} \left(\partial_ {\theta}I\right)^{2} +\frac{1}{r^{2} \sin^{2}\theta}\left(\partial_{\phi}I\right)^{2}=0$. Substituting Eq. (\ref{eq:SB sep}) into Eq. (\ref{eq:SB I}), we get the equation for the radial part as

\begin{equation}
C_{4}\left(\partial_{r}W\right)^{4}+C_{2}\left(\partial_{r}W\right)^{2}+C_{0}=0,
\end{equation}

\noindent where

\noindent
\begin{eqnarray}
C_{4} & = & -2\beta g{}^{2},\nonumber \\
C_{2} & = & g\left(1-4\beta m^{2}\right),\nonumber \\
C_{0} & = & m^{2}-2\beta m^{4}-\frac{\omega^{2}}{f}.
\end{eqnarray}

\noindent There are four roots for the above equation, but only two roots have the physical meaning. Solving the equation and finishing the integration at the event horizon yield

\noindent
\begin{eqnarray}
W_{\pm} & = & \pm\int dr\frac{1}{\sqrt{fg}}\sqrt{\omega^{2}-m^{2}f+2\beta m^{4}f}\left[1+\beta\left(m^{2}+\frac{\omega^{2}}{f}\right)\right]\nonumber \\
 & = & \pm i\pi2M\omega\left(1+\frac{1}{2}\beta\left(m^{2}+4\omega^{2}\right)\right)+\left(\mathrm{real\: part}\right).
\label{1-0}
\end{eqnarray}

\noindent where  $f=g=1-\frac{2M}{r}$, and $+/-$ represent the outgoing/ingoing solutions. In section (2.3), the expression of the tunneling rate is not invariant under canonical transformations. To consider an invariance under canonical transformations, we follow the recent work \cite{APAS1,APAS2,APAS3} and adopt the expression $\oint p_r dr$. Thus the tunneling rate is

\begin{eqnarray}
\Gamma &\propto & exp[- Im\oint p_r dr] =exp\left[- Im\left(\int p_r^{out}dr-\int p_r^{in}dr\right)\right]\nonumber\\
&=& exp\left[\pm 2Im\int p_r^{out,in}dr\right].
\label{1-1}
\end{eqnarray}

\noindent The expression of $p_r$ in the above equation relates to $W(r)$ as $p_r = \partial_r W$. The solutions of $Im\int p_r^{out,in}dr$ were gotten in Eq. (\ref{1-0}), which is

\begin{eqnarray}
Im\oint p_r dr = 2Im W^{out}=4 \pi M\omega\left(1+\frac{1}{2}\beta \left(m^{2}+4\omega^{2}\right)\right).
\label{1-2}
\end{eqnarray}

\noindent If we use Eq. (\ref{1-2}) to calculate the tunneling rate, we will obtain two times Hawking temperature \cite{AAPS1,AAPS2,AAPS3}. This result is inconsistent with the standard Hawking temperature. After careful considerations, Akhmedova et al. found that the contribution of the temporal part of the action had been ignored in the calculation \cite{APAS1,APAS2,APAS3}. When this contribution is considered, the problem that there is a factor of two in the Hawking temperature would be resolved.

To find the contributions of the temporal part, we introduce the Kruskal coordinates $(T,R)$. In this coordinate system, the exterior region of the black hole $(r>r_+)$ is described by

\begin{eqnarray}
T &=& e^{\kappa_+r_*}sinh(\kappa_+t),\nonumber\\
R &=& e^{\kappa_+r_*}cosh(\kappa_+t),
\label{eq1-3}
\end{eqnarray}

\noindent where $r_*= r+ \frac{1}{2\kappa_+}ln\frac{r-r_+}{r_+}$ is the tortoise coordinate, and $\kappa_+ =\frac{1}{2r_{+}}$ is the surface gravity at the horizon. The interior region is given by

\begin{eqnarray}
T &=& e^{\kappa_+r_*}cosh(\kappa_+t),\nonumber\\
R &=& e^{\kappa_+r_*}sinh(\kappa_+t).
\label{eq1-4}
\end{eqnarray}

\noindent To connect these two regions across the horizon, we need to rotate the time $t$ to $t\rightarrow t-i\frac{\pi}{2\kappa_+}$. As pointed in Ref. \cite{APAS1}, this "rotation" gives us an additional imaginary contribution of the temporal part, namely, $Im(\omega\Delta t^{out,in})=\frac{\pi \omega}{2\kappa_+}$. Then the total temporal contribution is $Im(\omega\Delta t)=\frac{\pi \omega}{\kappa_+}$.  Therefore, the tunneling rate of the scalar particle crossing the horizon is

\begin{eqnarray}
\Gamma &\propto & exp\left[-\left(Im (E\Delta t)+Im\oint p_r dr\right)\right] \nonumber\\
& = & \exp\left[-8\pi M\omega\left(1+\frac{1}{4}\beta \left(m^{2}+4\omega^{2}\right)\right)\right].
\label{eq1-5}
\end{eqnarray}

\noindent This is the Boltzmann factor with the Hawking temperature taking

\begin{eqnarray}
T = \frac{1}{8\pi M\left[1+\frac{1}{4}\beta\left(m^{2}+4\omega^{2}\right)\right]}
 = \left[1-\frac{1}{4}\beta\left(m^{2}+4\omega^{2}\right)\right]T_{0},
\label{eq1-5}
\end{eqnarray}

\noindent where $T_0 = \frac{1}{8\pi M}$ is the original Hawking temperature. Clearly, when quantum gravity effects are taken into account, there is a small correction to the Hawking temperature and the correction is not only affected by the quantum numbers (the mass and energy) of the emitted particle but also determined by the mass of the black hole. The corrected temperature is lower than the original one, which implies that quantum gravity correction slows down the increase of the Hawking temperature.

\subsection{Tunneling radiation of fermions}

\subsubsection{The generalized Dirac equation}
We study the tunneling radiation of fermions with effects of quantum gravity in this subsection. The fermion's equation of motion obeys Dirac equation. Therefore, we should first modify Dirac equation by the influence of quantum gravity effects. There are several ways to modify it.

The first way is to generalize GUP into the covariant form \cite{Kober:2010sj}. The modification of operators of positions and momenta are given as follows

\begin{eqnarray}
\hat{x}^{\mu} & = & x^{\mu},\nonumber \\
\hat{p}_{\mu} & = & -i\left(1-\beta\partial^{\rho}\partial_{\rho}\right)\partial_{\mu},
\end{eqnarray}

\noindent where $\mu$, $\rho$ run from $0$ to $3$. Therefore, the modified Dirac equation is

\begin{equation}
i\gamma^{\mu}\left(1-\beta\partial^{\rho}\partial_{\rho}\right)\partial_{\mu}\psi+m\psi=0.
\end{equation}

\noindent To consider this condition carefully, we should emphasize that the generalized energy under this modification becomes $\tilde{E}=E\left(1-\beta \partial^{\rho} \partial_{\rho}\right)$, where $E=i\partial_{t}$.

In the second way, the invariance under gauge transformations is taken into account\cite{Kober:2010sj}. Thus, it leads us to modify the operator

\begin{equation}
\left(1-\beta\partial^{\rho}\partial_{\rho}\right)\partial_{\mu}\rightarrow\left(1-\beta D^{\rho}D_{\rho}\right)D_{\mu},
\end{equation}

\noindent where $D_{\mu}\equiv\partial_{\mu}-ieA_{\mu}$. Therefore, we can define the generalized covariant derivative as

\begin{eqnarray}
\mathcal{D}_{\mu}&\equiv&\left(1-\beta D^{\rho}D_{\rho}\right)D_{\mu}\nonumber\\
 &=& \left[1-\beta\left(\partial^{\rho}-ieA^{\rho}\right)\left(\partial_{\rho}-ieA_{\rho}\right)\right]\left(\partial_{\mu}-ieA_{\mu}\right).
\end{eqnarray}

\noindent Here the action of the Dirac field is invariant under global gauge transformations

\begin{equation}
i\beta\bar{\psi}D^{\rho}D_{\rho}\gamma^{\mu}D_{\mu}\psi\rightarrow i\beta\bar{\psi}D^{\rho}D_{\rho}\gamma^{\mu}D_{\mu}\psi,
\end{equation}

\noindent with

\begin{eqnarray}
\gamma^{\mu} & \rightarrow & U\left(x\right)\gamma^{\mu}U\left(x\right)^{\dagger},\quad
U\left(x\right)^{\dagger}U\left(x\right) =  1.
\end{eqnarray}

\noindent The generalized field strength tensor can be defined as

\begin{equation}
\mathcal{F}_{\mu\nu}=\frac{i}{e}\left[\mathcal{D}_{\mu},\mathcal{D}_{\nu}\right].
\end{equation}

\noindent Then the generalized Dirac equation is gotten as

\begin{equation}
i\gamma^{\mu}\mathcal{D}_{\mu}\psi+m\psi=0.
\end{equation}

In the third way, the energy frequency and momentum operator are modified, respectively \cite{Hossenfelder:2003jz}. It is given as follows. Let us recall the Dirac equation with an electromagnetic field in curved spacetime

\begin{eqnarray}
\gamma^{\mu}\left(\partial_{\mu}+\Omega_{\mu}+\frac{i}{\hbar}eA_{\mu}\right)\Psi+\frac{m}{\hbar}\Psi=0,
\label{eq4.1.4}
\end{eqnarray}

\noindent which can be rewritten as

\begin{eqnarray}
-\gamma^{0}\partial_{0}\Psi=\left(\gamma^{i}\partial_{i}+\gamma^{\mu}\Omega_{\mu}+\gamma^{\mu}
\frac{i}{\hbar}eA_{\mu}+\frac{m}{\hbar}\right)\Psi. \label{eq4.1.5}
\end{eqnarray}

\noindent To introduce the effects of quantum gravity, we substitute Eqs. (\ref{eq4.1.1}) and (\ref{eq4.1.3}) into Eq. (\ref{eq4.1.5}) and keep the leading orders of $\beta$. Then the generalized Dirac equation in curved spacetime is obtained as

\begin{eqnarray}
-\gamma^{0}\partial_{0}\Psi=\left(\gamma^{i}\partial_{i}+\gamma^{\mu}\Omega_{\mu}+\gamma^{\mu}
\frac{i}{\hbar}eA_{\mu}+\frac{m}{\hbar}\right)\left(1+\beta\hbar^{2}\partial_{j}\partial^{j}-\beta
m^{2}\right)\Psi, \label{eq4.1.6}
\end{eqnarray}

\noindent namely,

\begin{eqnarray}
\left[\gamma^{0}\partial_{0}+\gamma^{i}\partial_{i}\left(1-\beta
m^{2}\right)+\gamma^{i}\beta\hbar^{2}\left(\partial_{j}\partial^{j}\right)\partial_{i}+\frac{m}{\hbar}
\left(1+\beta\hbar^{2}\partial_{j}\partial^{j}-\beta m^{2}\right)\right.\nonumber \\
\left.+\gamma^{\mu}\frac{i}{\hbar}eA_{\mu}\left(1+\beta\hbar^{2}\partial_{j}\partial^{j}-\beta
m^{2}\right)+\gamma^{\mu}\Omega_{\mu}\left(1+\beta\hbar^{2}\partial_{j}\partial^{j}-\beta
m^{2}\right)\right]\Psi= 0.  \label{eq4.1.7}
\end{eqnarray}

\noindent The above equation describes the motion of an uncharged fermion when $e= 0$. In the following subsections, we use the generalized Dirac equation (\ref{eq4.1.7}) to discuss the tunneling radiation of fermions with effects of quantum gravity.

\subsubsection{Fermions' tunneling in the Schwarzschild black hole}

We first review the fermion's tunneling across the horizon of the Schwarzschild black hole. The effects of quantum gravity are taken into account. The Schwarzschild metric is given by Eq. (\ref{eq2.1.1}). The gamma matrices and the wave function with spin up state are also given by Eqs. (\ref{eq2.3.6}) and (\ref{eq2.3.8}), respectively. To consider the quantum gravity effects, we substitute the matrices  (\ref{eq2.3.6}) and the wave function  (\ref{eq2.3.8}) into the generalized Dirac equation (\ref{eq4.1.7}). After the calculation, we find four decoupled equations

\begin{eqnarray}
 - iA\frac{1}{\sqrt f }\partial _t I - B\left( {1 - \beta m^2}
 \right)\sqrt g \partial _r I \nonumber \\
 - iAm\beta \left[ {g^{rr}\left( {\partial _r I} \right)^2 + g^{\theta \theta }\left( {\partial _\theta I} \right)^2
+ g^{\phi \phi }\left( {\partial _\phi I} \right)^2} \right] \nonumber \\
 + B\beta \sqrt g \partial _r I \left[ {g^{rr}\left( {\partial _r I}
\right)^2 + g^{\theta \theta }\left( {\partial _\theta I}
\right)^2 + g^{\phi \phi }\left( {\partial _\phi I} \right)^2}
\right] + iAm\left( {1 - \beta m^2} \right)  =  0, \label{eq4.2.1}
\end{eqnarray}

\begin{eqnarray}
iB\frac{1}{\sqrt f }\partial _t I - A\left( {1 - \beta m^2}
\right)\sqrt g\partial _r I \nonumber \\
- iBm\beta \left[ {g^{rr}\left( {\partial _r I} \right)^2 +
g^{\theta \theta }\left( {\partial _\theta I} \right)^2 + g^{\phi
\phi }\left( {\partial _\phi I} \right)^2} \right] \nonumber \\
 + A\beta \sqrt g \partial _r I \left[ {g^{rr}\left( {\partial _r I}
\right)^2 + g^{\theta \theta }\left( {\partial _\theta I} \right)^2
+ g^{\phi \phi }\left( {\partial _\phi I} \right)^2} \right] + iBm\left( {1 -
\beta m^2} \right) =  0,
\label{eq4.2.2}
\end{eqnarray}

\begin{eqnarray}
A\{-i(1-\beta m^2) \sqrt {g^{\phi \phi}}\partial _ {\phi}I \nonumber \\
+ i\beta\sqrt {g^{\phi \phi}}\partial _ {\phi}I[g^{rr} (\partial _r I)^2 +
g^{\theta \theta}(\partial _ {\theta}I)^2 + g^{\phi \phi}(\partial
_ {\phi}I)^2]\nonumber\\
- \sqrt {g^{\theta \theta}}\partial _ {\theta} I[1-\beta m^2 - \beta g^{rr}
(\partial _r I)^2 -\beta g^{\theta \theta}(\partial _ {\theta}I)^2
-\beta g^{\phi \phi}(\partial _ {\phi}I)^2 ]\} = 0, \label{eq4.2.3}
\end{eqnarray}

\begin{eqnarray}
B\{-i(1-\beta m^2) \sqrt {g^{\phi \phi}}\partial _ {\phi}I \nonumber \\
+ i\beta\sqrt {g^{\phi \phi}}\partial _ {\phi}I[g^{rr} (\partial _r I)^2 +
g^{\theta \theta}(\partial _ {\theta}I)^2 + g^{\phi \phi}(\partial
_ {\phi}I)^2]\nonumber\\
- \sqrt {g^{\theta \theta}}\partial _ {\theta} I[1-\beta m^2 - \beta g^{rr}
(\partial _r I)^2 -\beta g^{\theta \theta}(\partial _ {\theta}I)^2
-\beta g^{\phi \phi}(\partial _ {\phi}I)^2 ]\}  = 0. \label{eq4.2.4}
\end{eqnarray}

\noindent $\sqrt {g^{\theta \theta}}= \frac {1}{r}$,  $\sqrt {g^{\phi \phi}} = \frac {1} {r \sin\theta}$ in the above equations. The contributions from $\partial A$, $\partial B$ and the high orders of $\hbar$ were neglected. Moreover, since the metric has a time-like killing vector, the action $I$ can be also separated by Eq. (\ref{eq2.3.13}). Now, we insert Eq. (\ref{eq2.3.13}) into Eqs. (\ref{eq4.2.3}) and (\ref{eq4.2.4}) and divide by $A$, $B$. Then, Eqs. (\ref{eq4.2.3}) and (\ref{eq4.2.4}) are reduced to the following equation

\begin{eqnarray}
(\sqrt {g^{\theta \theta}}\partial _ {\theta} J +i\sqrt
{g^{\phi \phi}}\partial _ {\phi}J)\nonumber \\
\times [\beta g^{rr}(\partial _r W)^2 + \beta g^{\theta
\theta}(\partial _ {\theta}J)^2 +\beta g^{\phi \phi}(\partial
_ {\phi}J)^2 + \beta m^2 - 1] =0. \label{eq4.2.5}
\end{eqnarray}

\noindent Since $\beta$ is a small quantity represented the effects from quantum gravity, the value in the square bracket can not be vanished. Therefore, the above equation  is simplified as

\begin{eqnarray}
\sqrt {g^{\theta \theta}}\partial _ {\theta} J +i\sqrt
{g^{\phi  \phi}}\partial _ {\phi}J =0. \label{eq4.2.6}
\end{eqnarray}

\noindent As it is discussed in section (2.3), although the solution of $J$ is the complex functions, its contribution would be canceled out in the calculation of the tunneling rate. Therefore, we can neglect it. Meanwhile, the square of Eq. (\ref{eq4.2.6}) yields $g^{\theta\theta}\left(\partial_{\theta}J\right)^{2} + g^{\phi\phi}\left(\partial_{\phi}J\right)^{2} = 0$. Now, we focus our attention to the radial part of the action. Inserting Eq. (\ref{eq2.3.13}) into Eqs. (\ref{eq4.2.1}) and (\ref{eq4.2.2}) and canceling $A$ and $B$ in these equations yield

\begin{equation}
 A_6\left( {\partial _r W} \right)^6 + A_4\left(
{\partial _r W} \right)^4 + A_2\left( {\partial _r W} \right)^2 + A_0 =
0,
\label{eq4.2.7}
\end{equation}

\noindent where
\begin{eqnarray}
A_6 & = & \beta^{2}g^{3}f,\nonumber \\
A_4 & = & \beta g^{2}f\left(3m^{2}\beta -2\right),\nonumber \\
A_2 & = & gf\left[\left(1-\beta m^{2}\right)^{2}-\beta\left(2m^{2}-
2m^{4}\beta \right)\right],\nonumber \\
A_0 & = & m^{2}\left(1-\beta m^{2}\right)^{2}f-
\omega^{2}.
\label{eq4.2.8}
\end{eqnarray}

\noindent Neglecting the higher orders of $\beta$ and solving the above equations at the event horizon, we get

\begin{eqnarray}
W_{\pm} & = & \pm\int dr\frac{1}{\sqrt{fg}}\sqrt{\omega^{2}-m^{2}f+2\beta m^{4}f}\left[1+\beta\left(m^{2}+\frac{\omega^{2}}{f}\right)\right]\nonumber \\
 & = & \pm i\pi2M\omega\left(1+\frac{1}{2}\beta\left(m^{2}+4\omega^{2}\right)\right)+\left(\mathrm{real\: part}\right).
\label{eq4.2.9}
\end{eqnarray}

\noindent In the above equation, $+ / -$ signs are correspondent with outgoing/ingoing solutions and the real part is irrelevant to the tunneling rate \cite{Mitra}. To consider an invariance under canonical transformations, we follow the process in the above section \cite{APAS1} and derive the temporal contribution in the Kruskal coordinate system. Finally, the tunneling rate of the spin-1/2 fermion across the horizon is derived as

\begin{eqnarray}
\Gamma &\propto & exp\left[-\left(Im (E\Delta t)+Im\oint p_r dr\right)\right] \nonumber\\
& = & \exp\left[-8\pi M\omega\left(1+\frac{1}{4}\beta \left(m^{2}+4\omega^{2}\right)\right)\right],
\label{eq4.2.14}
\end{eqnarray}

\noindent which shows that the Hawking temperature is

\begin{eqnarray}
T = \frac{1}{8\pi M\left[1+\frac{1}{4}\beta\left(m^{2}+4\omega^{2}\right)\right]}
 = \left[1-\frac{1}{4}\beta\left(m^{2}+4\omega^{2}\right)\right]\frac{1}{8\pi M}.
\label{eq4.2.15}
\end{eqnarray}

\noindent Clearly, the corrected temperature is lower than the original one and the correction value is affected by the quantum numbers (the mass and energy) of the emitted fermion. The quantum correction slows down the increase of the Hawking temperature caused by the evaporation.

\subsubsection{Fermions' tunneling in the Reissner-Nordstrom spacetime}
Now let us discuss the effect of an electromagnetic field on the tunneling radiation of an charged fermion with effects of quantum gravity. Here, we use the Reissner-Nordstrom black hole which describes a spherically symmetric static spacetime with a charge $Q$. The Reissner-Nordstrom metric is given by

\begin{eqnarray}
ds^{2}=-f\left(r\right)dt^{2}+g\left(r\right)^{-1}dr^{2}+r^{2}\left(d\theta^{2}+\sin^{2}\theta
d\phi^{2}\right), \label{eq4.3.1}
\end{eqnarray}

\noindent with the electromagnetic potential

\begin{eqnarray}
A_{\mu}=\left(A_{t},0,0,0\right)= \left(\frac{Q}{r},0,0,0\right),
\label{eq4.3.2}
\end{eqnarray}

\noindent where $f\left(r\right)=g\left(r\right)=1-\frac{2M}{r}+\frac{Q^{2}}{r^{2}}= \frac{(r-r_+)(r-r_-)}{r^2}$, $r_\pm = M \pm \sqrt{M^2-Q^2}$ are the locations of the outer/inner horizons. We also only consider the case of the spin up state. The discussion of the spin down is parallel. The wave function with spin up state is

\begin{eqnarray}
\Psi=\left(\begin{array}{c}
A\\
0\\
B\\
0
\end{array}\right)\exp\left(\frac{i}{\hbar}I\left(t,r,\theta,\phi\right)\right),
\label{eq4.3.4}
\end{eqnarray}

\noindent where $I$ is the action and $A$, $B$ are functions of $t, r , \theta , \phi$. For the metric (\ref{eq4.3.1}), one can easily construct the tetrad as $ e_{\mu}^{\quad a}=\mathrm{diag}\left(\sqrt{f},1/\sqrt{g},r,r\sin\theta\right)$. Then, the $\gamma^\mu$ matrices are given by

\begin{eqnarray}
\gamma^{t}=\frac{1}{\sqrt{f\left(r\right)}}\left(\begin{array}{cc}
i & 0\\
0 & -i
\end{array}\right), &  & \gamma^{\theta}=\sqrt{g^{\theta\theta}}\left(\begin{array}{cc}
0 & \sigma^{1}\\
\sigma^{1} & 0
\end{array}\right),\nonumber \\
\gamma^{r}=\sqrt{g\left(r\right)}\left(\begin{array}{cc}
0 & \sigma^{3}\\
\sigma^{3} & 0
\end{array}\right), &  & \gamma^{\phi}=\sqrt{g^{\phi\phi}}\left(\begin{array}{cc}
0 & \sigma^{2}\\
\sigma^{2} & 0
\end{array}\right),
\label{eq4.3.5}
\end{eqnarray}

\noindent where $\sqrt{g^{\theta\theta}} =\frac{1}{r}$ and $\sqrt{g^{\phi\phi}} =\frac{1}{r\sin\theta}$. To apply the WKB approximation, we substitute the wave
function and gamma matrices into the generalized Dirac equation. Multiplying $\hbar$ and considering the leading contribution, we get the equations of motion as follows

\begin{eqnarray}
-iA\frac{1}{\sqrt{f}}\partial_{t}I +iAm\left[1-\beta m^{2}
-\beta g^{rr}\left(\partial_{r}I\right)^{2}-\beta g^{\theta\theta}\left(\partial_{\theta}I\right)^{2}-\beta
g^{\phi\phi}\left(\partial_{\phi}I\right)^{2}\right]\nonumber\\
-B\sqrt{g}\partial_{r}I
\left[1-\beta m^{2}
-\beta g^{rr}\left(\partial_{r}I\right)^{2}-\beta g^{\theta\theta}\left(\partial_{\theta}I\right)^{2}-\beta
g^{\phi\phi}\left(\partial_{\phi}I\right)^{2}\right]\nonumber\\
-iA\frac{eA_t}{\sqrt f}\left[1-\beta m^{2}
-\beta g^{rr}\left(\partial_{r}I\right)^{2}-\beta g^{\theta\theta}\left(\partial_{\theta}I\right)^{2}-\beta
g^{\phi\phi}\left(\partial_{\phi}I\right)^{2}\right] = 0,
\label{eq4.3.6}
\end{eqnarray}

\begin{eqnarray}
iB\frac{1}{\sqrt{f}}\partial_{t}I+iBm\left[1-\beta m^{2}
-\beta g^{rr}\left(\partial_{r}I\right)^{2}-\beta g^{\theta\theta}\left(\partial_{\theta}I\right)^{2}-\beta
g^{\phi\phi}\left(\partial_{\phi}I\right)^{2}\right]\nonumber\\
-A\sqrt{g}\partial_{r}I
\left[1-\beta m^{2}
-\beta g^{rr}\left(\partial_{r}I\right)^{2}-\beta g^{\theta\theta}\left(\partial_{\theta}I\right)^{2}-\beta
g^{\phi\phi}\left(\partial_{\phi}I\right)^{2}\right]\nonumber\\
+iB\frac{eA_t}{\sqrt f}\left[1-\beta m^{2}
-\beta g^{rr}\left(\partial_{r}I\right)^{2}-\beta g^{\theta\theta}\left(\partial_{\theta}I\right)^{2}-\beta
g^{\phi\phi}\left(\partial_{\phi}I\right)^{2}\right]= 0,
\label{eq4.3.7}
\end{eqnarray}

\begin{eqnarray}
A\left\{-(1-\beta m^2) \sqrt {g^{\theta \theta}}\partial _
{\theta} I
+ \beta \sqrt {g^{\theta \theta}}\partial _
{\theta}I\left[g^{rr}(\partial _r I)^2 + g^{\theta
\theta}(\partial _ {\theta}I)^2
+ g^{\phi \phi}(\partial _ {\phi}I)^2 \right]\right.\nonumber\\
\left.-i \sqrt {g^{\phi \phi}}\partial _ {\phi}I
\left[1-\beta m^{2}
-\beta g^{rr}\left(\partial_{r}I\right)^{2}-\beta g^{\theta\theta}\left(\partial_{\theta}I\right)^{2}-\beta
g^{\phi\phi}\left(\partial_{\phi}I\right)^{2}\right]\right\} = 0,
\label{eq4.3.8}
\end{eqnarray}

\begin{eqnarray}
B\left\{-(1-\beta m^2) \sqrt {g^{\theta \theta}}\partial _
{\theta} I
+ \beta \sqrt {g^{\theta \theta}}\partial _
{\theta}I\left[g^{rr}(\partial _r I)^2 + g^{\theta
\theta}(\partial _ {\theta}I)^2
+ g^{\phi \phi}(\partial _ {\phi}I)^2 \right]\right.\nonumber\\
\left.-i \sqrt {g^{\phi \phi}}\partial _ {\phi}I
\left[1-\beta m^{2}
-\beta g^{rr}\left(\partial_{r}I\right)^{2}-\beta g^{\theta\theta}\left(\partial_{\theta}I\right)^{2}-\beta
g^{\phi\phi}\left(\partial_{\phi}I\right)^{2}\right]\right\} = 0.
\label{eq4.3.9}
\end{eqnarray}

\noindent However, it is difficult to solve the action from the above equations directly. We should follow the standard process to carry out separation of variables. The action becomes

\begin{eqnarray}
I = -\omega t + W(r) + \Theta (\theta , \phi). \label{eq4.3.10}
\end{eqnarray}

\noindent We first observe the last two equations in (\ref{eq4.3.6})-(\ref{eq4.3.9}). Inserting Eq. (\ref{eq4.3.10}) into Eqs. (\ref{eq4.3.8}) and (\ref{eq4.3.9}) and canceling $A$ and $B$, we find the equation $\sqrt {g^{\theta \theta}}\partial _ {\theta} \Theta +i\sqrt {g^{\phi \phi}}\partial _ {\phi}\Theta =0$. It finally yields an important relation: $g^{\theta \theta}(\partial _ {\theta}I)^2 + g^{\phi \phi}(\partial _ {\phi}I)^2 = 0$. Next, we process to consider Eqs. (\ref{eq4.3.6}) and (\ref{eq4.3.7}). Inserting Eq. (\ref{eq4.3.10}) into Eqs. (\ref{eq4.3.6}) and (\ref{eq4.3.7}) and canceling $A$ and $B$ yield

\begin{eqnarray}
A_6\left(\partial_{r}W\right)^{6}+A_4\left(\partial_{r}W\right)^{4}+A_2\left(\partial_{r}W\right)^{2}+A_0=0,
\label{eq4.3.11}
\end{eqnarray}

\noindent where

\begin{eqnarray}
A_6 &=& \beta^{2}g^{3}f,\nonumber\\
A_4 &=& \beta g^{2}f\left(3m^{2}\beta-2\right)-\beta ^2 g^2 e^2A_t^2,\nonumber\\
A_2 &=& gf\left[\left(1-\beta m^{2}\right)^{2}-2\beta m^{2}\left(1-\beta m^{2}\right)\right]+2\beta g eA_t [-\omega + eA_t(1-\beta m^2)],\nonumber\\
A_0 &=& m^{2}f\left(1-\beta m^{2}\right)^{2}-\left[-\omega
+eA_t\left(1-\beta m^{2}\right)\right]^2. \label{eq4.3.12}
\end{eqnarray}

\noindent  The particle's tunneling rate is determined by the imaginary part of the action. Neglecting higher order terms of $\beta$, the imaginary part of the radial action is solved as follows

\begin{eqnarray}
Im W_\pm (r) & = & \pm Im \int dr\frac{1}{\sqrt{gf}}\sqrt{-m^{2}\left(1-2\beta m^{2}\right)f+\left(\omega-eA_{t}\left(1-\beta m^{2}\right)\right)^{2}}\nonumber \\
&& \times \left\{1+\beta\left[m^{2}-\frac{eA_{t}\tilde{\omega}_{0}}{f}+\frac{1}{f}\left(\omega-eA_{t}\left(1-\beta m^{2}\right)\right)^{2}\right]\right\} \nonumber \\
& = & \pm \pi\frac{r_+^2}{r_+ - r_-} (\omega - eA_{t+})
\times\left(1+ \beta\Xi\right) ,
 \label{eq4.3.13}
\end{eqnarray}

\noindent where $+(-)$ represent the outgoing (ingoing) solutions, $\tilde\omega_0 = \omega - eA_{t}$ and $A_{t+} = \frac{Q}{r_+}$ is the electromagnetic potential at the event horizon. $\Xi$ is given by

\begin{eqnarray}
\Xi & = &\frac{m^2}{2}+\frac{em^2A_{t+}}{\omega - eA_{t+}} \nonumber \\
&&+ 2\frac{4e\omega Qr_+r_- + \omega^2 r_+^3 -2e^2Q^2(r_+ + r_-) -e\omega Q r_+^2 - 2\omega^2r_+^2r_-}{\left(r_+-r_-\right)^2r_+}.
\label{eq4.3.14}
\end{eqnarray}

\noindent Utilizing $r_{\pm} = M \pm \sqrt{M^2-Q^2}$, the right hand side of Eq. (\ref{eq4.3.14}) is reduced to $\frac{m^2}{2}+\frac{em^2A_{t+}}{\omega - eA_{t+}}+ 2\frac{2eQ^2\left(\omega Q -eM \right) + \omega r_+ \left(\omega - eA_{t+}\right)\left(r_+^2-Q^2\right)}{\left(r_+-r_-\right)^2r_+}$. It is easy to find $\Xi >0$.

We use the invariant expression of the tunneling rate under canonical transformations. Therefore, we should first find the contributions of the temporal part. Introducing the tortoise coordinate $r_*= r+ \frac{1}{2\kappa_+}ln\frac{r-r_+}{r_+}- \frac{1}{2\kappa_-}ln\frac{r-r_-}{r_-}$, where $\kappa_\pm = \frac{ r_+-r_- } {2r_{\pm}^2}$ is the surface gravity at the outer (inner) horizons, we get the exterior region of the black hole in the Kruskal coordinates system, which is

\begin{eqnarray}
T = e^{\kappa_+r_*}sinh(\kappa_+t), \quad
R = e^{\kappa_+r_*}cosh(\kappa_+t),
\label{eq4.3.15}
\end{eqnarray}

\noindent The interior region is described as

\begin{eqnarray}
T = e^{\kappa_+r_*}cosh(\kappa_+t), \quad
R = e^{\kappa_+r_*}sinh(\kappa_+t).
\label{eq4.3.16}
\end{eqnarray}

\noindent We rotate the time $t$ to $t\rightarrow t-\frac{i\pi}{2\kappa_+}$ to connect these two regions. This rotation produces an additional imaginary contribution, namely, $Im(E\Delta t^{out,in})=\frac{\pi E}{2\kappa_+}$, where $E=\omega - e A_{t+}$. Considering that the total temporal contribution is $Im(E\Delta t)=\frac{\pi E}{\kappa_+}$, we get the expression of the tunneling rate as follows

\begin{eqnarray}
\Gamma &\propto & exp\left[-\left(Im (E\Delta t^{out,in})+Im\oint p_r dr\right)\right] \nonumber\\
&=& exp\left[-\left(Im (E\Delta t)+ 2 Im W^{out}\right)\right] \nonumber\\
& = & \exp\left[ -4\pi\frac{r_+^2 }{r_+ - r_-}(\omega - e A_{t+}) \left(1 + \frac{1}{2}\beta\Xi\right)\right].
\label{eq4.3.17}
\end{eqnarray}

\noindent This is the Boltzmann factor for an object with the effective temperature taking

\begin{eqnarray}
T = \frac{r_+ - r_-}{4\pi r_+^2\left(1+
\frac{1}{2}\beta\Xi\right)}= T_0\left(1 - \frac{1}{2}\beta\Xi\right) ,
\label{eq4.3.18}
\end{eqnarray}

\noindent where $T_0 =  \frac{r_+ - r_-}{4\pi r_+^2}$ is the original Hawking temperature of the Reissner-Nordstrom black hole. It is easily found that the corrected temperature is lower than the original one. The correction value relies on the quantum numbers (charge, mass, energy) of the emitted charged fermion. The quantum correction also slows down the increase of the Hawking temperature. When $\beta = 0$, the original Hawking temperature of the Reissner-Nordstrom black hole is recovered.

\subsubsection{Fermions' tunneling in the 5-dimensional rotating black string}
The Kerr metric describes a rotating black hole in four dimensions. When the Kerr metric is added to an extra compact spatial dimension, the metric takes a form

\begin{eqnarray}
ds^{2}& =& -\frac{\Delta}{\rho^2}\left(dt - a\sin^2\theta d\varphi\right)^2+ \frac{\sin^2\theta}{\rho^2}\left[adt - (r^2+a^2)d\varphi\right]^2 \nonumber\\
&&+\frac{\rho^2}{\Delta}dr^2 +\rho^2d\theta^2+ g_{zz}dz^2,
\label{eq4.4.1}
\end{eqnarray}

\noindent where $\Delta= r^2-2Mr +a^2 = (r-r_+)(r-r_-)$, $\rho^2  = r^2+a^2\cos^2\theta$, $g_{zz}$ is set to be $1$. The metric describes a rotating uniform black string. $r_{\pm}= M\pm \sqrt{M^2-a^2}$ locates at the outer (inner) horizons. $M$ and $a$ are the mass and the angular momentum unit mass of the string. The equation of motion of a fermion is determined by the generalized Dirac equation. For the investigation of the tunneling behavior of the fermion, we first construct a tetrad and it can be directly constructed from the metric (\ref{eq4.4.1}). However, due to the rotation of the spacetime, the matter near the horizons is dragged by the spacetime (\ref{eq4.4.1}), which is inconvenient for us to discuss the fermion's tunneling behavior. To eliminate this inconvenience, we carry out the dragging coordinate transformation

\begin{eqnarray}
d\phi &=&  d\varphi - \Omega dt, \nonumber\\
\Omega &=& \frac{\left(r^2+a^2- \Delta\right) a}{\left( r^2+a^2\right)^2 -\Delta a^2\sin^2\theta},
\label{eq4.4.2}
\end{eqnarray}

\noindent on the metric (\ref{eq4.4.1}). Then the metric (\ref{eq4.4.1}) becomes

\begin{eqnarray}
ds^2 &=& - F(r)dt^2+\frac{1}{G(r)}dr^{2} +g_{\theta \theta}d\theta^2 +g_{\phi\phi} d\phi^2 + g_{zz}dz^2\nonumber\\
&=& - \frac{\Delta \rho^2}{(r^2+a^2)^2-\Delta a^2 \sin^2{\theta}}dt^2 +\frac{\rho^{2}}{\Delta}dr^{2} + g_{zz}dz^2+\rho^2 d\theta^2\nonumber\\
&& + \frac{\sin^2{\theta}}{\rho^2}\left[(r^2+a^2)^2-\Delta a^2 \sin^2{\theta}\right]d\phi^2.
\label{eq4.4.3}
\end{eqnarray}

\noindent From the above metric, we construct the tetrad as follows

\begin{eqnarray}
e_{\mu}^a= diag(\sqrt{F},1/\sqrt{G},\sqrt{g_{\theta\theta}},\sqrt{g_{\phi\phi}},\sqrt{g_{zz}}).
\label{eq4.4.4}
\end{eqnarray}

\noindent Now it is easy to construct the following gamma matrices

\begin{eqnarray}
\gamma^{t}=\frac{1}{\sqrt{F}}\left(\begin{array}{cc}
0 & I\\
-I & 0
\end{array}\right), &  & \gamma^{\theta}=\sqrt{g^{\theta\theta}}\left(\begin{array}{cc}
0 & \sigma^{2}\\
\sigma^{2} & 0
\end{array}\right),\nonumber \\
\gamma^{r}=\sqrt{G}\left(\begin{array}{cc}
0 & \sigma^{3}\\
\sigma^{3} & 0
\end{array}\right), &  & \gamma^{\phi}=\sqrt{g^{\phi\phi}}\left(\begin{array}{cc}
0 & \sigma^{1}\\
\sigma^{1} & 0
\end{array}\right),\nonumber \\
\gamma^{z}=\sqrt{g^{zz}}\left(\begin{array}{cc}
-I & 0\\
0 & I
\end{array}\right).
\label{eq4.4.5}
\end{eqnarray}

\noindent There are also two states with spin up and spin down to measure the quantum property of the spin-1/2 fermion. We only investigate the tunneling radiation of the fermion with the spin up state. The analysis of spin down case is parallel. The wave function of the fermion with spin up state in the spacetime (\ref{eq4.4.3}) are assumed as follows

\begin{eqnarray}
\Psi =\left(\begin{array}{c}
A\\
0\\
B\\
0
\end{array}\right)\exp\left(\frac{i}{\hbar}I\left(t,r,\theta,\phi,z\right)\right),
\label{eq4.4.6}
\end{eqnarray}

\noindent where $A, B, C, D$ are functions of $(t, r, \theta, \phi, z)$, and $I$ is the action of the fermion. To use the WKB approximation, we adopt the same process as the above section and get four equations

\begin{eqnarray}
-\frac{B}{\sqrt{F}}\partial_t I- B \sqrt{G}(1-\beta m^2)\partial_r I + A \sqrt{g^{zz}}(1-\beta m^2)\partial_z I \nonumber\\
- iA m(1-\beta m^2- \beta Q)+ B\beta \sqrt{G} Q \partial_r I - A\beta \sqrt{g^{zz}} Q \partial_z I = 0,
\label{eq4.4.8}
\end{eqnarray}

\begin{eqnarray}
\frac{A}{\sqrt{F}}\partial_t I- A \sqrt{G}(1-\beta m^2)\partial_r I -B \sqrt{g^{zz}}(1-\beta m^2)\partial_z I \nonumber\\
- iB m(1-\beta m^2- \beta Q)+ A\beta \sqrt{G} Q \partial_r I +B\beta \sqrt{g^{zz}} Q \partial_z I = 0,
\label{eq4.4.9}
\end{eqnarray}

\begin{eqnarray}
-B\left(i\sqrt{g^{\theta\theta}}\partial_{\theta} I + \sqrt{g^{\phi\phi}}\partial_{\phi} I \right)(1-\beta m^2 - \beta Q) = 0,
\label{eq4.4.10}
\end{eqnarray}

\begin{eqnarray}
-A\left(i\sqrt{g^{\theta\theta}}\partial_{\theta} I + \sqrt{g^{\phi\phi}}\partial_{\phi} I \right)(1-\beta m^2 - \beta Q) = 0,
\label{eq4.4.11}
\end{eqnarray}

\noindent where $Q= g^{rr} \left( {\partial _r I} \right)^2  + g^{\theta \theta } \left( {\partial _\theta  I} \right)^2 + g^{\phi \phi } \left( {\partial _\phi  I} \right)^2+ g^{zz} \left( {\partial_z I} \right)^2$. Following the same process, we carry out separation of variables

\begin{eqnarray}
I = -(\omega-j\Omega)t + W(r) + \Theta (\theta,\phi) +Jz,
\label{eq4.4.12}
\end{eqnarray}

\noindent where $\omega$ and $j$ are  the energy and angular momentum of the fermion, and $J$ is a conserved momentum corresponding to the compact dimension. From Eqs. (\ref{eq4.4.10}) and (\ref{eq4.4.11}), we can get a complex function solution of $\Theta$, and

\begin{eqnarray}
g^{\theta\theta}(\partial_{\theta} \Theta)^2 + g^{\phi\phi}(\partial_{\phi} \Theta)^2 = 0.
\label{eq4.4.14}
\end{eqnarray}

\noindent Now we focus our attention on the radial part of the action. Insert Eq. (\ref{eq4.4.12}) into Eqs. (\ref{eq4.4.8}) and (\ref{eq4.4.9}), cancel $A$ and $B$ and neglect the higher order terms of $\beta$. We get the solution of $\partial_rW$

\begin{eqnarray}
A(\partial_r W)^4 + B(\partial_r W)^2 + C = 0,
\label{eq4.4.15}
\end{eqnarray}

\noindent where

\begin{eqnarray}
A &=& -2\beta G^2F,\nonumber\\
B &=& G,\nonumber\\
C &=& -\left(1-2\beta m^{2}-2\beta g^{zz}\left(\partial_{z}I\right)^{2}\right)\left(m^{2}+g^{zz}\left(\partial_{z}I\right)^{2}\right)-\frac{1}{F}\left(\partial_{t}I\right)^{2}.
\label{eq4.4.16}
\end{eqnarray}

\noindent Solving the above equation at the outer horizon yields the complex expression of the radial action. However, the tunneling rate is determined by the imaginary part of the action. Therefore, the imaginary part of $W$ is derived as

\begin{eqnarray}
Im W_{\pm} &=&  Im\int dr\frac{1}{\sqrt{GF}}\sqrt{\left(E-j\Omega\right)^{2}+\left(1-2\beta m^{2}-2\beta g^{zz}J^{2}\right)\left(m^{2}+g^{zz}J^{2}\right)F}\nonumber\\
&& \times \left[1+\beta\left(m^{2}+g^{zz}J^{2}+\frac{\left(E-j\Omega\right)^{2}}{F}\right)\right]\nonumber\\
&=& \pi \frac{(\omega -j\Omega_+)(r_+^2 +a^2)}{r_+ - r_-}\left[1+\beta \Xi(J,\theta,r_+,j)\right],
\label{eq4.4.18}
\end{eqnarray}

\noindent where $g^{zz} =1$ and $\Omega_+ = \frac{a}{r_+^2 +a^2}$ is the angular velocity at the outer horizon. $\Xi(J,\theta,r_+,j) $ is a complicated function of $J,\theta,r_+,j$ and  $\Xi(J,\theta,r_+,j) >0$. To find the contributions of the temporal part, we should use the Kruskal coordinates $(T,R)$. In this coordinate system, the exterior region $(r>r_+)$ of the Kerr string is described by

\begin{eqnarray}
T &=& e^{\kappa_+r_*}sinh(\kappa_+t),\nonumber\\
R &=& e^{\kappa_+r_*}cosh(\kappa_+t),
\label{eq4.4.19}
\end{eqnarray}

\noindent where $r_*= r+ \frac{1}{2\kappa_+}ln\frac{r-r_+}{r_+}- \frac{1}{2\kappa_-}ln\frac{r-r_-}{r_-}$, and $\kappa_\pm =\frac{r_+ - r_-}{2(r_{\pm}^2+a^2)}$. The interior region of the string is given by

\begin{eqnarray}
T &=& e^{\kappa_+r_*}cosh(\kappa_+t),\nonumber\\
R &=& e^{\kappa_+r_*}sinh(\kappa_+t).
\label{eq4.4.20}
\end{eqnarray}

\noindent For connecting these two regions across the horizon, we need to rotate the time $t$ to $t\rightarrow t-i\frac{\pi}{2\kappa_+}$. This ��rotation�� gives us  an additional imaginary contribution of the temporal part \cite{APAS1}, namely, $Im(E\Delta t^{out,in})=\frac{\pi E}{2\kappa_+}$, where $E=\omega - j \Omega_+$. Then the total temporal contribution is $Im(E\Delta t)=\frac{\pi E}{\kappa_+}$.  When all of the contributions are taken into account, the tunneling rate of the fermion across the horizon is derived as follows

\begin{eqnarray}
\Gamma &\propto & exp\left[-\left(Im (E\Delta t)+Im\oint p_r dr\right)\right] \nonumber\\
&=& -4\pi \frac{(\omega -j\Omega_+)(r_+^2 +a^2)}{ r_+ - r_- }\left[1+\frac{1}{2}\beta \Xi(J,\theta,r_+,j)\right].
\label{eq4.4.21}
\end{eqnarray}

\noindent This is the Boltzman factor corresponding to the Hawking temperature of the Kerr string

\begin{eqnarray}
T &=& \frac{ r_+ - r_- }{4\pi (r_+^2 +a^2)\left[1+\frac{1}{2}\beta \Xi(J,\theta,r_+,j)\right]}\nonumber\\
&=& T_0\left[1-\frac{1}{2}\beta \Xi(J,\theta,r_+,j)\right],
\label{eq4.4.22}
\end{eqnarray}

\noindent where $T_0=  \frac{ r_+ - r_- }{4\pi (r_+^2 +a^2)}$ is the original Hawking temperature of the Kerr string, which shares the same expression as that of the 4-dimensional Kerr black hole. This modified temperature is not only determined by the mass, angular momentum and extra dimension of the string, but also affected by the quantum numbers (energy, mass and angular momentum) of the fermion. Meanwhile, a different phenomenon is that the temperature is affected by the angular $\theta$. Due to the small $\beta$, the quantum gravity correction to the temperature is also a small value.

The metric (\ref{eq4.4.1}) reduces to the Schwarzschild string metric when $a=0$. Then the imaginary part of the radial action (\ref{eq4.4.18}) reduces to

\begin{eqnarray}
Im\oint p_r dr&=& 2 \pi \omega r_+\left[1+\beta \left(2\omega^2 +m^2/2+J^2/2\right)\right].
\label{eq4.4.23}
\end{eqnarray}

\noindent Adopting the same process, the temperature of the Schwarzschild string is gotten

\begin{eqnarray}
T = \frac{1}{8\pi M}\left[1-\beta \left(\omega^2 +m^2/4+J^2/4\right)\right].
\label{eq4.4.24}
\end{eqnarray}

\noindent It shows that the temperature of the string is affected by the effect of the extra dimension and the quantum number (energy, mass and angular momentum) of the fermion. It is also easy to see that the quantum gravity correction slows down the increase of the temperature.

\subsection{Remnants in black holes}

Incorporating effects of quantum gravity, the tunneling radiation of scalar particles and fermions were discussed in the above subsections. The corrected temperatures are lower than the original temperatures and the correction values are related to the quantum numbers of the emitted particles. This result implies that quantum gravity corrections slow down the increases of the temperatures during the radiation. Finally, the quantum corrections and the traditional temperatures rising tendency are in balance. The radiation should be stopped at a certain point, which naturally leads to the existence of the remnants in the black holes.

It is of interest to discuss the scale of the remnants. When $Q=0$ and $a=0$, $J=0$, the metrics of the Reissner-Nordstrom black hole and Kerr string reduce to that of the Schwarzschild black hole.  Then Eqs. (\ref{eq4.2.15}), (\ref{eq4.3.18}) and (\ref{eq4.4.22}) share an identical expression and show the same temperature. This temperature is

\begin{eqnarray}
T = \left[1-\frac{1}{4}\beta\left(m^{2}+4\omega^{2}\right)\right]\frac{1}{8\pi M},
\label{eq4.5.1}
\end{eqnarray}

\noindent To determine the scale of the remnants, we only consider the tunneling radiation of a massless particle. When

\begin{equation}
(M-dM)(1+\beta \omega^2)\simeq M,
\label{eq4.5.2}
\end{equation}

\noindent the radiation stops,  where $dM = \omega$, $\beta =\beta_0/M_p^2$, $\beta_0 <10^{5}$ is a dimensionless parameter determined by quantum gravity effects \cite{LMS}. To avoid the temperature $T$ being negative, the value of $\omega$ should satisfy $1-\beta\omega^2>0$, namely, $\omega < \frac{M_p} {\sqrt{\beta_0}}$. Then, it is easy to find the remnant and the corresponding temperature

\begin{equation}
M_{\hbox{Res}} \simeq \frac{M_p^2}{\beta_0 \omega} \gtrsim
\frac{M_p}{\sqrt{\beta_0}}.
\label{final results}
\end{equation}

This result is consistent with those derived in references \cite{ACS,BG,CA,XIANG}. To compare with previous works, our process presents how the remnants of black holes arises by quantum gravity effects. Moreover, the singularity of black hole evaporation is also prevented by the quantum gravity correction.

\section{The mass spectrum}

Quantizing the geometry is an important topic of quantum effects of black holes. The horizon area was first quantized by Bekenstein \cite{JDB}. The area spectrum satisfies the formula $A_n = n\gamma l_p^2$, where $n =1,2,3...$, $\gamma$ is a dimensionless constant and needs to be fixed and $l_p$ is the Planck length. Later Vaz et al. examined WDW equation for the Schwarzschild black hole and found that the mass is also quantized and the mass spectrum is proportional to $\sqrt{n}$ \cite{VW,V}. In this section, we first review the quantization of a Schwarzschild black hole by WDW equation, and then take into account quantum gravity effects to re-quantize the black hole \cite{MR,OST}.

To get the energy of the Schwarzschild black hole, one can write down the time independent Schrodinger equation $\hat{H} \psi = E \psi$ of the black hole, where  $\hat{H}$ is the Hamiltonian operator and $\psi$ is the wave function of the black hole. $E_n$ are the eigenvalues of $\hat{H}$. The eigenfunctions $\psi_n$ describe the corresponding probability amplitudes. WDW equation for the Schwarzschild black hole where the Hamiltonian involves momenta and coordinates $(a, p_a)$ i.e. $ H = \frac{p_a^2}{2a} +\frac{a}{2}$, can be written as \cite{MR,OST}

\begin{eqnarray}
a^{-s-1}\frac{\partial}{\partial a}\left(a^s\frac{\partial}{\partial a}\right) \psi (a)=(a-r_E)\psi (a),
\label{1.24}
\end{eqnarray}

\noindent where $p_a$ is momenta canonically conjugate to $a$, $p_a^2=-a^{-s}\frac{d}{da}\left(a^s\frac{d}{da}\psi (a)\right)$, $r_E= 2M$ is the location of the horizon and $s$ is a factor ordering the parameter. To solve the above equation, we choose a particular value $s = 2$. Thus, the above equation is rewritten as follows

\begin{eqnarray}
\frac{1}{a}\left(\frac{\partial^2}{\partial a^2} +\frac{2}{a}\frac{\partial}{\partial a}\right) \psi (a)=(a-r_E)\psi (a).
\label{1.25}
\end{eqnarray}

\noindent We first perform the transformations $\psi (a)=\frac{U(a)}{a}$ and $\xi = a -r_E$, where $\xi$ expresses the gravitational degrees of freedom of the black hole. Then define the appropriate constants and consider the fact that the energy of excitations associated with variable $a$ is not positive. Thus, Eq. (\ref{1.25}) becomes

\begin{eqnarray}
-\frac{\partial^2 U}{\partial \xi^2} +\xi^2U=\frac{1}{4}r_E^2U,
\label{1.26}
\end{eqnarray}

\noindent which is the differential equation for a quantum harmonic oscillator. Its energy levels are easily derived as

\begin{eqnarray}
\frac{1}{4}r_E^2 =2n+1,
\label{1.27}
\end{eqnarray}

\noindent where $n$ is an non-negative integer. Using the relation between the radius and mass, we get the mass spectrum of the black hole as

\begin{eqnarray}
M^2 =2(n+1/2).
\label{1.28}
\end{eqnarray}

\noindent From the above equation, we can find that the mass of the black hole is proportional to $\sqrt{n}$. This result is full in consistence with that of Bekenstein's argument \cite{JDB2,JDB3}.

In the following, we investigate the mass spectrum by the modified WDW equation with effects of quantum gravity. WDW equation for a Schwarzschild black hole with a modified Heisenberg algebra which has a linear term in momentum was discussed in Ref.\cite{BM111}. Here we adopt the expression of the GUP put forward by Mann et al. \cite{KMM}. The position and momentum operators in position space were given in Eq. (\ref{eq1.2}). They are represented in momentum space as follows

\begin{eqnarray}
X &=&i\hbar\left[(1+\beta p^2)\partial p+\beta p\right],\nonumber\\
P&=&p ~.
\label{1.29}
\end{eqnarray}

\noindent Then WDW equation for the Schwarzschild black hole with the GUP becomes \cite{BJM}

\begin{eqnarray}
\left\{-\left[(1+\beta p^2)\partial p+\beta p\right]^2-\beta (1+\beta p^2) +p^2\right\} \psi(p) = \frac{r_E^2}{4}\psi(p),
\label{1.30}
\end{eqnarray}

\noindent where $p$ is canonical momenta conjugate to $\xi$. The exact solution of the above equation was found in the work \cite{CMOT,DJM}, which is

\begin{eqnarray}
M^2 = 2\left(n+1/2\right)\sqrt{1+(\beta /2)^2}+\beta (n^2+n+1/2).
\label{1.31}
\end{eqnarray}

\noindent The normalized energy eigenfunctions of Eq. (\ref{1.30}) are gotten as

\begin{eqnarray}
\psi_n(p)=2^{\nu}\Gamma \sqrt{\frac{n!(n+\nu)\sigma ^{1/2}}{2\pi \Gamma (n+2\nu)}}c^{\nu+1}C_n^{\nu}(s),
\label{1.32}
\end{eqnarray}

\noindent where

\begin{eqnarray}
C_n^{\nu}(s) = \frac{(-1)^n}{2^nn!}\frac{\Gamma (n+2\nu)\Gamma (\nu +1/2)}{\Gamma (2\nu)\Gamma (n+\nu +1/2)}(1-s^2)^{1/2-\nu}\frac{d^n}{ds^n}(1-s^2)^{n+\nu-1/2}
\label{eq1.29}
\end{eqnarray}

\noindent is the Gegenbauer polynomial, $n=0,1,2,....$, $c=(1+\beta p^2)^{-1/2}$, $s=\sqrt{\beta}pc$ and $2\nu = 1+\sqrt{1+4/{\beta ^2}}$. Comparing Eq. (\ref{1.31}) with Eq. (\ref{1.28}), we find that the correction value appeared when effects of quantum gravity are taken into account. The mass of the black hole is proportional to $n$ in quantum gravity regime. When $\beta = 0$, Eq. (\ref{1.31}) reduces to Eq. (\ref{1.28}) and Bekenstein's result is recovered \cite{JDB2,JDB3}.

\section{Thermodynamics}

In this section, we discuss the thermodynamics of the Schwarzschild black hole with effects of quantum gravity. The first law of thermodynamics of a black hole satisfies $dM = TdS -VdQ -\Omega dJ$. The entropy is one quarter of the horizon area, namely, $S= A/4$. To derive the entropy, we can directly solve Eq. (\ref{1.4}) which is the modified Heisenberg algebra. Neglecting the factor "2" on the right hand side of the equation and solving it, we get the expression of $\Delta p$ as follows

\begin{eqnarray}
\Delta p = \frac{2M/(M_p^2c)+\alpha_1  \pm \sqrt{\left(2M/(M_p^2c)+\alpha_1  \right)^2-4\alpha_2 }}{2\alpha_2 }.
\label{1.13}
\end{eqnarray}

\noindent In the above equation, the negative sign is chosen to get the temperature. The positive sign has no evident physical meaning. Then we use the calibration factor $4 \pi$ and get the temperature as follows

\begin{eqnarray}
T= \frac{c\Delta p}{4\pi} = \frac{2M/M_p^2+\alpha_1 c  }{8\pi\alpha_2 }\left(1-\sqrt{1-\frac{4\alpha_2}{\left(2M/(M_p^2c)+\alpha_1  \right)^2}}\right).
\label{1.14}
\end{eqnarray}

\noindent When $\alpha_1 =0$, $\alpha_2=\beta$, we can get the correction to the temperature derived by Adler et al.  \cite{ACS}. Using the first law of thermodynamics $dS= c^2dM/T$, we finish the integral and get the expression of the entropy as follows

\begin{eqnarray}
S &=& 2\pi M^2/M_p^2 + 2\alpha_1 \pi Mc \nonumber\\
&&-2\pi\alpha_2 M_p^2c^2 ln\left({\alpha_1 M_p^2 c +2M+\sqrt{4M^2  +4\alpha_1 M M_p^2c +\alpha_1^2 M_p^4c^2 -4\alpha_2 M_p^4c^2}}\right)\nonumber\\
&& + \pi\left(2M/M_p^2+\alpha_1 c/2\right) \sqrt{4M^2  +4\alpha_1 M M_p^2c +\alpha_1^2 M_p^4c^2 -4\alpha_2 M_p^4c^2}.
\label{1.15}
\end{eqnarray}

\noindent The above entropy was derived by the DSR-GUP. A logarithmic correction to the entropy was recently derived in the theories of loop quantum gravity and string \cite{KM}. There is also a logarithmic correction term. When $\alpha_1= \alpha_2 =0$, the above equation is reduced to the Bekenstein-Hawking entropy of the Schwarzschild black hole.

Now we derive the thermodynamics from another aspect \cite{MAJ,LPB,CAP}. We rewrite Eq. (\ref{1.4}) as follows

\begin{eqnarray}
\Delta x \Delta p \geq \hbar\left(1 - \alpha_1 \Delta p+ \alpha_2 \Delta p^2\right),
\label{1.16}
\end{eqnarray}

\noindent where $\alpha_1 = \sqrt{\alpha_2}= \alpha_0 l_p /{\hbar}$, $\alpha_0$ is a dimensionless constant. Solving the above equation and Taylor expanding the solved equation, we rewrite the same equation after some simple manipulation as follows

\begin{eqnarray}
\Delta p = \frac{1}{\delta x}\left(1 - \frac{\alpha_1}{2\delta x} +\frac{\alpha_1^2}{2(\delta x)^2}-\frac{\alpha_1^3}{2(\delta x)^3}+\frac{9\alpha_1^2}{16(\delta x)^2}-...\right).
\label{1.17}
\end{eqnarray}

\noindent In the above equation, we have chosen $\hbar = 1$. In the work \cite{MAJ,LPB,CAP}, the authors discussed that the Heisenberg uncertainty principle could be translated to the lower bound $E\delta x \geq 1$, where $E$ is the energy of a quantum particle. Here we adopt this consideration and take into account a photon as the quantum particle. Then the lower bound is rebuilt by the GUP, which is

\begin{eqnarray}
E = \frac{1}{\delta x}\left(1 - \frac{\alpha_1}{2\delta x} +\frac{\alpha_1^2}{2(\delta x)^2}-\frac{\alpha_1^3}{2(\delta x)^3}+\frac{9\alpha_1^2}{16(\delta x)^2}-...\right).
\label{1.18}
\end{eqnarray}

\noindent When a classical particle with energy $E$ and size $R$ is absorbed by the black hole, the minimal increase in the area is expressed as

\begin{eqnarray}
\Delta A_{min} \geq 8\pi l_p^2 ER\geq 8\pi l_p^2 E\delta x ,
\label{1.19}
\end{eqnarray}

\noindent Now we follow the work of Majumder and rewrite the above equation, which is \cite{MAJ}

\begin{eqnarray}
\Delta A_{min} \simeq \epsilon l_p^2\left(1 - \frac{\alpha_1}{2\delta x} +\frac{\alpha_1^2}{2(\delta x)^2}-\frac{\alpha_1^3}{2(\delta x)^3}+\frac{9\alpha_1^2}{16(\delta x)^2}- ...\right),
\label{1.20}
\end{eqnarray}

\noindent where $\epsilon$ is a numerical value greater than the order of $8\pi$. Let the uncertainty of the position be $\delta x \sim 2r_E = \sqrt{\frac{A}{\pi}}$, we get

\begin{eqnarray}
\Delta A_{min} \simeq \epsilon l_p^2\left(1 - \frac{\pi ^{1/2}\alpha_1}{A^{1/2}} +\frac{\pi\alpha_1^2}{A}-\frac{\pi^{3/2}\alpha_1^3}{A^{3/2}}+\frac{9\pi^2\alpha_1^2}{16A^2}- ...\right).
\label{1.21}
\end{eqnarray}

\noindent When the area is added, the entropy should be also added. We assume that the minimal increase of the entropy is one $bit~ b$ and the value of $b$ is $ln 2$. Thus, there is

\begin{eqnarray}
\frac{dS}{dA}\simeq\frac{\Delta S_{min}}{\Delta A_{min}}=\frac{b}{ \epsilon l_p^2\left(1 - \frac{\pi ^{1/2}\alpha_1}{A^{1/2}} +\frac{\pi\alpha_1^2}{A}-\frac{\pi^{3/2}\alpha_1^3}{A^{3/2}}+\frac{9\pi^2\alpha_1^2}{16A^2}-...\right)}.
\label{1.22}
\end{eqnarray}

\noindent The entropy can be obtained by integrating the above equation. Finishing the integral and letting $b/{\epsilon} = 1/4$, we get

\begin{eqnarray}
S &=& \frac{A}{4l^2_p}+\frac{\pi^{1/2}\alpha_0}{2}\sqrt{\frac{A}{4l^2_p}}\nonumber\\
&&-\frac{\pi \alpha_0^2}{16}ln{\frac{A}{4l^2_p}} -\frac{\pi^{3/2}\alpha_0^3}{32}\left(\frac{A}{4l^2_p}\right)^{-1/2} +\frac{\pi^2 \alpha_0^4}{128}\left(\frac{A}{4l^2_p}\right)^{-1} -....
\label{1.23}
\end{eqnarray}

\noindent It shows that the entropy is corrected by quantum gravity effects. The first term is the Bekenstein-Hawking entropy-area relation. The logarithmic correction appears in the expression, which is in consistence with that obtained in loop quantum gravity. For the Reissner-Nordstrom black hole with double horizons, the uncertainty of position is chosen as $\delta x = 2(r_+-r_-)$. Following the above process, we can also get the corrected entropy.

\section{Discussion and Conclusion}

In this paper, we have reviewed the effects of quantum gravity on the black holes. We first discussed the tunneling radiation. In the derivation of the tunneling radiation, the calculation of the imaginary part of the emitted particle's action is a key point. There are two ways to derive the imaginary part, namely, the null geodesic method and the Hamilton-Jacobi method. In the first way, the background spacetimes were varied, therefore, the leading corrections to the Hawking temperatures appeared. The corrected temperatures were higher than the original temperatures. In the second way, the spacetimes were fixed, thus the standard Hawking temperatures were recovered. Both of these results predict the complete evaporation of the black holes.

We can directly derive the action of the massive and massless particles in the Hamilton-Jacobi method. However, we should distinguish the massless and massive particles in null geodesic method, since the derivations of their actions are different. Taking into account that the derivation of the geodesic equation of the massive particle was gotten unnaturally, we re-defined the geodesic equation by the variation principle on the Lagrangian action. Then we adopted the new definition to investigate the tunneling behavior of the scalar particles across the horizon of the black string. The result is full in consistence with that derived by Parikh and Zhang et al. \cite{PW,ZZ1}.

When effects of quantum gravity were taken into account, we modified Klein-Gordon equation and Dirac equation by the influence of GUP, then used these modified equations to discuss the tunneling radiation of scalar particles and fermions in various spacetimes. The result showed that the corrected temperatures are lower than the original temperatures. Quantum corrections slow down the increase of the temperatures duo to the radiation. Finally, at a certain point, the quantum corrections and the traditional temperatures rising tendency are in balance, which leads to the remnants. The scale of the remnants was gotten by the massless particles' tunneling across the horizons, which is  $M_{\hbox{Res}}\gtrsim \frac{M_p}{\sqrt{\beta_0}}$. Then the singularities are avoided by the quantum gravity corrections. Our calculation explicitly showed how the remnants arises due to effects of quantum gravity. The tunneling radiation of scalar particles and fermions beyond the semi-classical approximation was also discussed. One can refer to Refs. \cite{BM1,BM2,SVZ} for detail.

Incorporating the quantum gravity effects into WDW equation, we quantized the Schwarzschild black hole's mass. The result showed that the mass of the black hole is proportional to $n$ in quantum gravity regime. When $\beta =0$, the result of Bekenstein is recovered. The thermodynamics of the Schwarzschild black hole were discussed with the influences of GUP. The corrected entropy was derived. There is also a logarithmic correction term, which is full in consistence with that derived in loop quantum gravity.

\section*{Acknowledgments}

This work is supported by the NSFC (Grant Nos. 11205125, 11175039, 11375121) and Sichuan Province Science Foundation for Youths (Grant Nos. 2014JQ0040, 2012JQ0039 ).

\end{document}